\documentclass{aa}  

\usepackage{graphicx}
\usepackage{txfonts}
\usepackage{tikz}
\usepackage{hyperref}

\usepackage{subcaption}
\begin{document} 

\newcommand{\solar}{$_\odot$}
\newcommand{\tento}[1]{$10^{#1}$}
\newcommand{\timestento}[2]{$#1 \times 10^{#2}$}
\newcommand{\MAa}{$M_{\rm Aa}$}
\newcommand{\MAbone}{$M_{\rm Ab1}$}
\newcommand{\MAbtwo}{$M_{\rm Ab2}$}
\newcommand{\aAaAb}{$a_{\rm Aab}$}
\newcommand{\aAbonebtwo}{$a_{\rm Ab1Ab2}$}
\newcommand{\RAa}{$R_{\rm Aa}$}
\newcommand{\RAbone}{$R_{\rm Ab1}$}
\newcommand{\RAbtwo}{$R_{\rm Ab2}$}
\newcommand{\iAaAb}{$i_{\rm AaAb}$}
\newcommand{\iAbonebtwo}{$i_{\rm Ab1Ab2}$}
\newcommand{\fAab}{$\phi_{\rm Aab}$}
\newcommand{\fAbonebtwo}{$\phi_{\rm Ab1b2}$}
\newcommand{\lfrac}{$f_{\rm Aa}/f_{\rm Total}$}

\newcommand{\M}{$M$}
\newcommand{\xc}{$X_{\rm c}$}
\newcommand{\Z}{$Z$}
\newcommand{\fov}{$f_{\rm ov}$}

\newcommand{\rotfreqomegasurf}{$\Omega_{\rm surf}$}
\newcommand{\rotfreqomegaenv}{$\Omega_{\rm env}$}
\newcommand{\rotfreqomegacore}{$\Omega_{\rm c}$}

\newcommand{\teff}{$T_{\rm eff}$}
\newcommand{\logg}{log$(g)$}
\newcommand{\lum}{$L$}
\newcommand{\logl}{log$(L)$}

\newcommand{\chisqr}{$\chi^2$}

\newcommand{\mone}{$^{-1}$}

\newcommand{\kepler}{\textit{Kepler}}
\newcommand{\starshadow}{\texttt{STAR SHADOW}}
\newcommand{\pofour}{\texttt{period04}}
\newcommand{\amigo}{\texttt{AMiGO}}
\newcommand{\cthree}{\texttt{C-3PO}}

\newcommand\kms{\ifmmode{\rm km\thinspace s^{-1}}
\else km\thinspace s$^{-1}$\fi}
\newcommand{\perd}{$\rm d^{-1}$}

\newcommand{\gssp}{\texttt{GSSP}}
\newcommand{\gdor}{$\gamma$~Dor}
\newcommand{\dsct}{$\delta$~Sct}
\newcommand{\patli}{\texttt{PAT\_LI2020}}
\newcommand{\patopt}{\texttt{PAT\_P04\_OPT}}
\newcommand{\patpes}{\texttt{PAT\_P04\_PES}}
\newcommand{\patsts}{\texttt{PAT\_STS}}

   \title{KIC~4150611: A quadruply eclipsing heptuple star system with a g-mode period-spacing pattern}

   \subtitle{Asteroseismic modelling of the g-mode period-spacing pattern}

   \author{
   Alex Kemp\inst{1}\thanks{
    \email{alex.kemp@kuleuven.be}}
    \and
    Dario J. Fritzewski \inst{1}
    \and
    Timothy Van Reeth\inst{1}
    \and
    Luc IJspeert\inst{1}
    \and
    Mathias Michielsen\inst{1}
    \and
    Joey S. G. Mombarg\inst{2}
    \and
    Vincent Vanlaer\inst{1}
    \and
    Gang~Li\inst{1}
    \and
    Andrew Tkachenko\inst{1}
    \and
    Conny Aerts\inst{1,3,4}
  }

   \institute{Institute of Astronomy (IvS), KU Leuven, Celestijnenlaan 200D, 3001, Leuven, Belgium
   \and
   IRAP, Universit\'{e} de Toulouse, CNRS, UPS, CNES, 14 Avenue Édouard Belin, 31400 Toulouse, France
   \and
   Department of Astrophysics, IMAPP, Radboud University Nijmegen, PO Box 9010, 6500 GL Nijmegen, The Netherlands
   \and
   Max Planck Institute for Astronomy, Koenigstuhl 17, 69117 Heidelberg, Germany}
   \date{}

 
  \abstract
   {KIC~4150611 is a high-order (seven) multiple composed of a triple system with: a F1V primary (Aa), which is eclipsed on a 94.2d period by a tight binary composed of two K/M dwarfs (Ab1, Ab2), which also eclipse each other; an eccentric, eclipsing binary composed of two G stars (Ba, Bb); and another faint eclipsing binary composed of two stars of unknown spectral type (Ca and Cb). In addition to its many eclipses, the system is an SB3 spectroscopic multiple (Aa, Ba, and Bb) and the primary (Aa) is a hybrid pulsator, exhibiting high amplitude pressure and gravity modes (g-modes). Further, its g-modes are arrayed in a period-spacing pattern, which greatly assists with mode identification and asteroseismic modelling. In aggregate, this richness in physics offers an excellent opportunity to obtain a precise physical characterisation for some of the stars in this system.}
   {In this work, we aim to estimate the stellar parameters of the primary (Aa) by performing asteroseismic analysis on its period-spacing pattern.}
   {We use the \cthree\ neural network to perform asteroseismic modelling of the g-mode period-spacing pattern of Aa, discussing the interplay of this information with external constraints from spectroscopy (\teff\ and \logg) and eclipse modelling ($R$). To estimate the level of uncertainty due to different frequency extraction and pattern identification processes, we consider four different variations on the period-spacing patterns. To better understand the correlations between and the uncertainty structure of our parameter estimates, we also employed a classical, parameter-based MCMC grid search on four different stellar grids.}
   {The best-fitting, externally constrained model to the 
   period-spacing pattern arrives at estimates of the stellar properties for Aa of: \M=$1.51\pm0.05$~M\solar, \xc=$0.43\pm0.04$, R=$1.66\pm0.1$~R\solar, \fov=0.010, \rotfreqomegacore=$1.58\pm0.01$~\perd with rigid rotation to within the measurement errors, log(\teff)=$3.856\pm0.008$~dex, \logg=$4.18\pm0.04$~dex, and \logl=$0.809\pm0.005$~dex, which agree well with previous measurements from eclipse modelling, spectroscopy, and the \emph{Gaia} DR3 luminosity.}
   {We find that
   the near-core properties of the best-fitting asteroseismic models are consistent with external constraints from eclipse modelling and spectroscopy. For stellar properties not relating to the near-core region, external constraints on the asteroseismic best-fitting models are informative. Aa appears to be a typical example of a \gdor\ star, fitting well within existing populations. We find that Aa is quasi-rigidly rotating to within the uncertainties, and note that the asteroseismic age estimate for Aa ($1100\pm100$~Myr) is considerably older than the young (35 Myr) age implied by previous isochrone fits to the B binary in the literature. Our MCMC parameter-based grid-search agrees well with our pattern-modelling approach. Improved future 
   modelling may come from 
   detailed coverage of metalicity effects and careful treatment of envelope physics.}


   \maketitle
%

\section{Introduction}

Asteroseismology, the study of stellar oscillations, stands as a cornerstone in modern astrophysics, offering unique insights into fundamental stellar properties. The sensitivity of stellar oscillation frequencies to the interior structure has opened the door to the measurement of stellar properties that are beyond the reach of surface observations. 
We can exploit this direct link between theoretical models and observations to refine our understanding of stellar evolution \citep{aerts2021}.


Asteroseismology's role in modern astrophysics has grown rapidly since the advent of space-based planet-hunting surveys such as CoRoT \citep{baglin2003}, \kepler\ \citep{borucki2010}, and TESS \citep{ricker2015}, which provided near-uninterrupted, high-precision photometry with regular cadence on long time-bases.
One outcome of this new era of space-based photometry has been the detection of low-frequency gravity-modes (g-modes) in large numbers of stars \citep{vanreeth2015method,vanreeth2015diagnostic,li2019rossby,li2019slow}. Previously, this had been impossible due to the logistical challenges of obtaining long time base data for even a small number of stars (e.g., \citealt{decat2002,aerts2004,decat2006}). Gravity-mode (g-mode) oscillations are waves that propagate with buoyancy as the dominant restoring force, and are particularly sensitive to the near-core stellar properties \citep{miglio2008}. Pressure modes (p-modes) have pressure as their dominant restoring force and have higher frequency. These modes are more sensitive to bulk stellar properties such as the average stellar density and envelope properties such as rotation \citep{aerts2010}.

In order for stellar oscillations to be observed, they must propagate to the surface where they produce variations in the stellar flux. Main sequence stars more massive than approximately 1.2 M\solar\ have a convective core and a radiative envelope. As g-modes are restored by buoyancy, they can propagate through the radiative envelope from the near-core region to the surface, where they are observed, but not within the convective stellar core.


Stars with masses between approximately 1.4~M\solar\ and 1.9~M\solar\ with observed g-mode pulsations are known as $\gamma$~Doradus (\gdor) pulsators, and feature g-mode pulsations excited via convective flux blocking \citep{dupret2005}. Some \gdor\ stars have overlap with the $\delta$ Scuti (\dsct) stars, which typically have masses between 1.5 and 2.5 M\solar, and exhibit p-mode oscillations driven by the opacity-driven heat engine mechanism.

The primary component of KIC~4150611\footnote{ HD~141469; RA=19$^{\circ}$ 18' 58.21759", DEC=+39$^{\circ}$ 16' 01.7913" (J2000)} is an example of a \gdor/\dsct\ hybrid pulsator: a star which exhibits both \dsct\ and \gdor\ pulsations \citep{uytterhoeven2011}. In this work, we focus our analysis attention on modelling the g-mode period-spacing pattern. The prominent p-mode pulsations do not form a part of this analysis, as without mode identification they provide negligible constraining power -- but add significant modelling complexity -- to the analysis. Analysis of the p-mode \dsct\ pulsations can be found in \cite{shibahashi2012} and \cite{balona2014}. 

\begin{figure}
\centering
\begin{tikzpicture}
\draw (0,1) -- (0,2);
\filldraw[black] (0,1) circle (2pt) node[anchor=north]{Aa};
\filldraw[white] (0,0.1) circle (2pt) node[anchor=south,color=black]{F1V};
\draw (0,2) -- (1,2) -- (1,2.2)  node[left]{A} node[right]{94.2d} -- (1,3);

\draw (1,2) -- (2,2) -- (2,1.7) node[left]{Ab} node[right]{1.52d} -- (2,1.5);

\draw (2,1.5) -- (1.5,1.5) -- (1.5,1);
\filldraw[black] (1.5,1) circle (2pt) node[anchor=north]{Ab1};
\filldraw[white] (1.5,0.1) circle (2pt) node[anchor=south,color=black]{K/M?};

\draw (2,1.5) -- (2.5,1.5) --(2.5,1.0);
\filldraw[black] (2.5,1) circle (2pt) node[anchor=north]{Ab2};
\filldraw[white] (2.5,0.1) circle (2pt) node[anchor=south,color=black]{K/M?};

\draw (1,2) -- (1,3);

\draw[dashed] (1,3) -- (2.5,3) node[anchor=south]{1000yr?} -- (4,3);
\draw[dotted] (4,3) -- (5,3) node[anchor=south]{?}-- (6,3);

\draw (4,3) -- (4,2) -- (4,2.2) node[left]{B} node[right]{8.65d} -- (4,2) --(3.5,2) -- (3.5,1.5);
\draw (3.8,2) -- (3.5,2);

\draw (3.5,2) -- (3.5,1.5);
\filldraw[black] (3.5,1.5) circle (2pt) node[anchor=north]{Ba};
\filldraw[white] (3.5,0.6) circle (2pt) node[anchor=south,color=black]{G};

\draw (4,2) -- (4.5,2) -- (4.5,1.5);
\filldraw[black] (4.5,1.5) circle (2pt) node[anchor=north]{Bb};
\filldraw[white] (4.5,0.6) circle (2pt) node[anchor=south,color=black]{G};

\draw (6,3) -- (6,2.2) node[left]{C} node[right]{1.43d} -- (6,2) -- (5.5,2);

\draw (5.5,2) -- (5.5,1.5);
\filldraw[black] (5.5,1.5) circle (2pt) node[anchor=north]{Ca};
\filldraw[white] (5.5,0.6) circle (2pt) node[anchor=south,color=black]{?};

\draw (6,2) -- (6.5,2) -- (6.5,1.5);
\filldraw[black] (6.5,1.5) circle (2pt) node[anchor=north]{Cb};
\filldraw[white] (6.5,0.6) circle (2pt) node[anchor=south,color=black]{?};

\end{tikzpicture}
\caption{Summarising the system hierarchy and nomenclature of KIC~4150611, with approximate orbital periods shown in days. \cite{heliminiak2017} present astrometric evidence for an association between the A and B components, but otherwise the A, B, and C components can be considered dynamically independent. Figure is based on Figure 15 from \protect \cite{heliminiak2017}.}
\label{fig:hierarchy}
\end{figure}
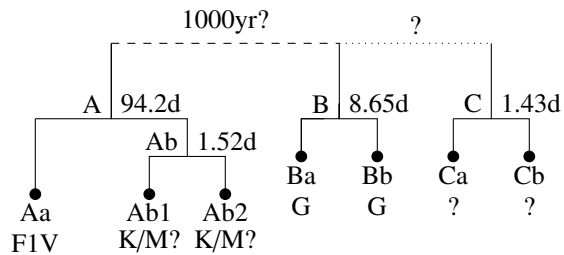

KIC~4150611 is a high-order (seven) multiple star system with four different sets of eclipses. The primary, Aa, is a bright ($V\approx8$ mag) F1V-type star that is in an eclipsing 94.2d circular orbit with a 1.52d self-eclipsing binary (Ab) composed of two dim (negligible light contribution) K/M dwarf stars (Ab1 and Ab2). The resulting primary eclipses of the triple geometry are complicated, varying significantly between eclipses depending on the phase of the Ab binary. Associated on roughly a 1000~yr orbit with the A triple is the eccentric 8.65d eclipsing B binary, composed of two near-identical G-type stars, Ba and Bb. The final component of the candidate heptuple is the C binary, a 1.43d eclipsing binary composed of two stars of unknown spectral type, Ca and Cb. If the C binary is indeed dynamically associated with the A-B quintuple -- and therefore at approximately the same distance from the observer -- their negligible contribution to the total flux implies that they are likely also cool dwarfs. The system structure is summarised in Fig. \ref{fig:hierarchy}.

The system structure described above was established in \cite{heliminiak2017}, who conducted  velocity modelling of Aa, Ba, and Bb as well as eclipse modelling of the Ab, B, and C binaries. The authors further performed imaging of the system using adaptive optics which established it as a visual triple, astrometric measurements searching for long-period dynamical associations, and isochrone fitting based on the properties of Ba and Bb. Detailed eclipse modelling of the A triple can be found in \cite{kemp2024eclipse}, along with spectroscopic analysis and atmospheric modelling of the disentangled spectra of Aa, Ba, and Bb.

In this work, we build on the previous eclipse and atmospheric analysis of Aa to perform asteroseismic modelling of its g-mode period-spacing pattern, first identified in \cite{li2020binary}. G-mode period-spacing patterns are sensitive to interior stellar properties such as near-core rotation rates and buoyancy travel times \citep{vanreeth2016,mombarg2019}, and when combined with grids of stellar models with computed pulsations, can be used to estimate stellar properties such as masses, ages, and mixing profiles \citep[see, for example,][]{aerts2018,johnston2019,pedersen2021,mombarg2021,michielsen2023}. By considering several different period-spacing patterns and by employing both a pattern-matching approach and parameter-based grid-search approach that leverages different grids of stellar models, we aim to provide insight into the systematic uncertainty associated with different methods and choices. Attention is also paid to how the different non-asteroseismic measurements that have been made for this system interplay with both each-other and the asteroseismology.

We provide a summary of physical constraints relevant to the asteroseismic modelling of KIC~4150611's g-mode period-spacing pattern in Section \ref{sec:lit_constraints}. We describe our methodology in Section \ref{sec:meth}, and present our results in Section \ref{sec:results}. We conclude in Section \ref{sec:conc}.

\section{Physical constraints from previous literature}

\begin{figure*}
\centering
   \includegraphics[width=1.99\columnwidth]{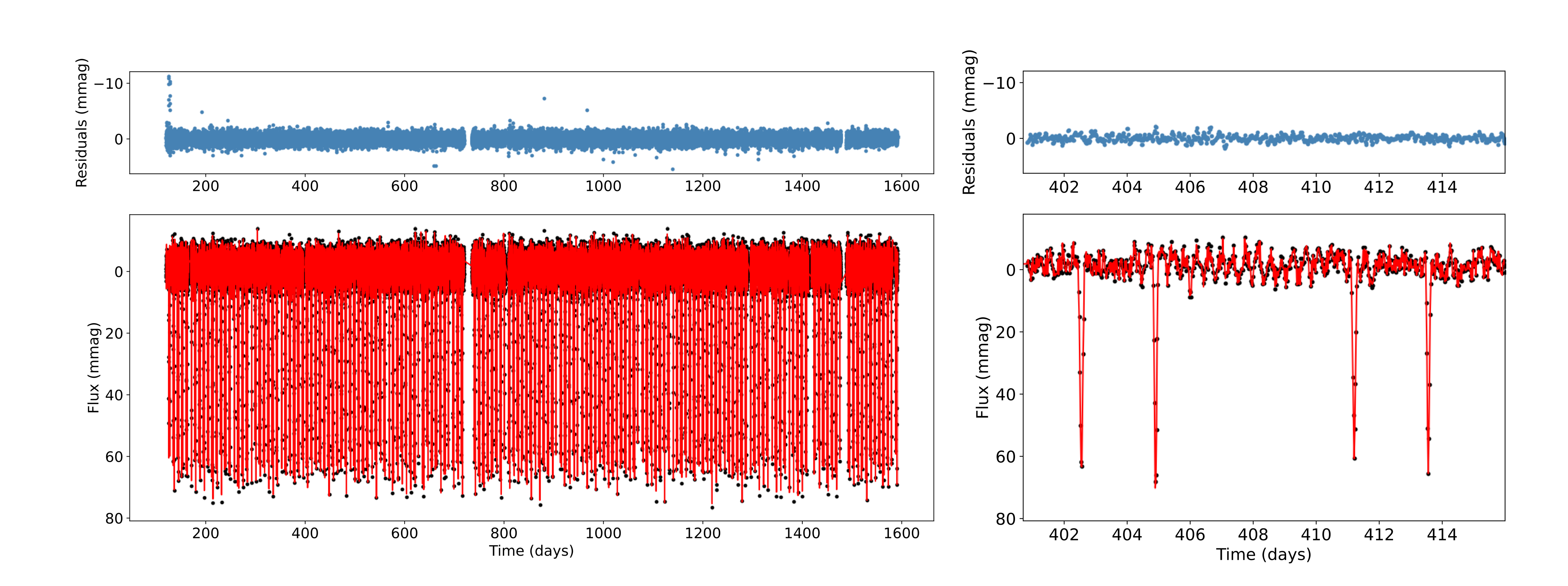}
       \caption{The bottom panel shows the normalised, detrended light curve (black) (excepting 94.2d eclipses) and the sinusoid model (red) formed from all frequencies extracted using \starshadow. The upper panel shows the residuals (blue). Figure 1 of \cite{kemp2024eclipse} shows the equivalent figure for the \pofour\ extraction.}
    \label{fig:lightcurve_sts}
\end{figure*}

\begin{figure}
\centering
   \includegraphics[width=0.99\columnwidth]{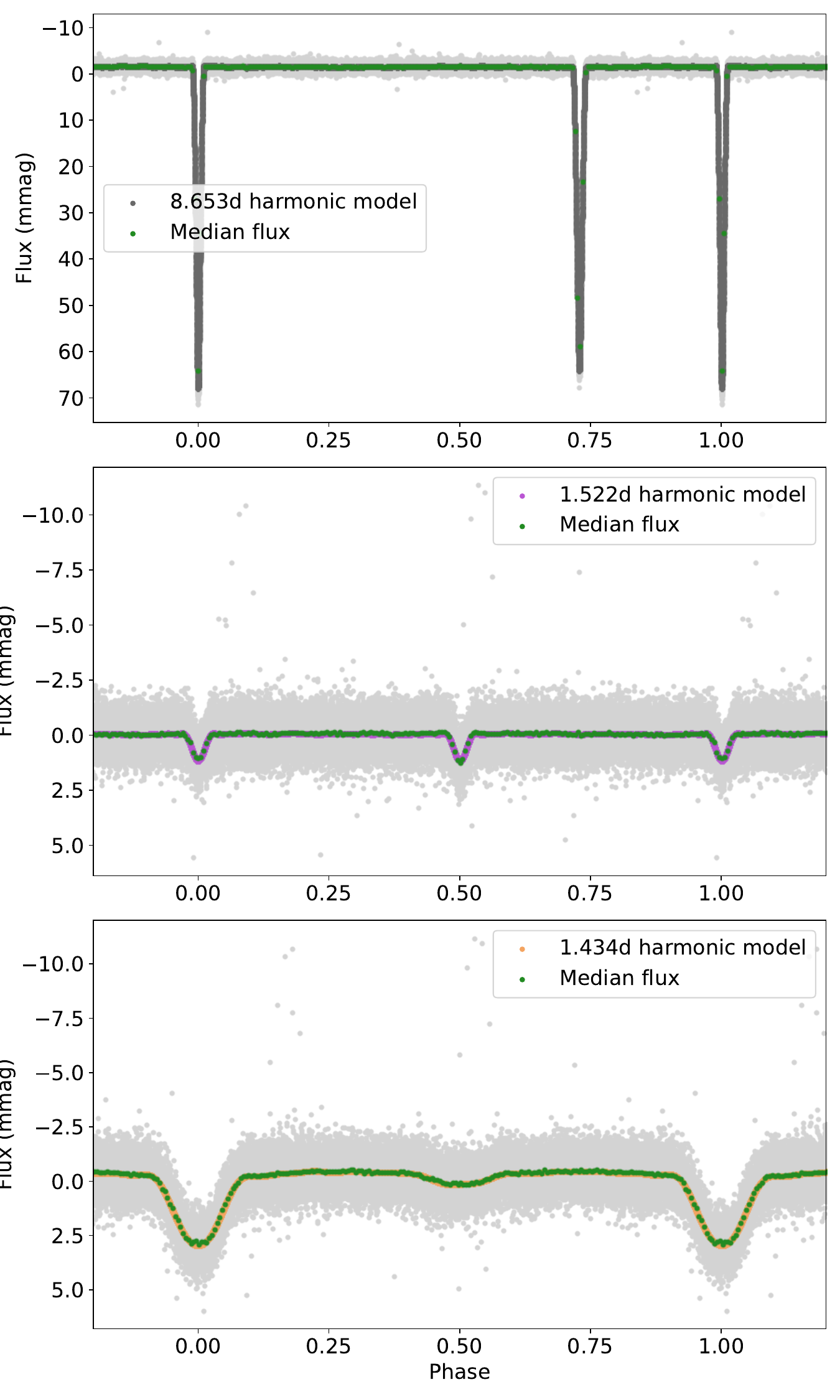}
       \caption{Phase-folded light curves for each of the eclipsing components and its sinusoid model. The light curve (grey data) for each is the residual between the normalised, detrended light curve and every frequency except those forming the sinusoid model of the relevant eclipse from \starshadow. The median flux is shown in green, while the coloured (colour varies by panel) points are from relevant harmonic model. Fig. 3 of \cite{kemp2024eclipse} shows the equivalent figure for the \pofour\ extraction, with no significant differences to the eclipse geometry.}
    \label{fig:phasefold_sts}
\end{figure}

\begin{figure*}
\centering
   \includegraphics[width=1.99\columnwidth]{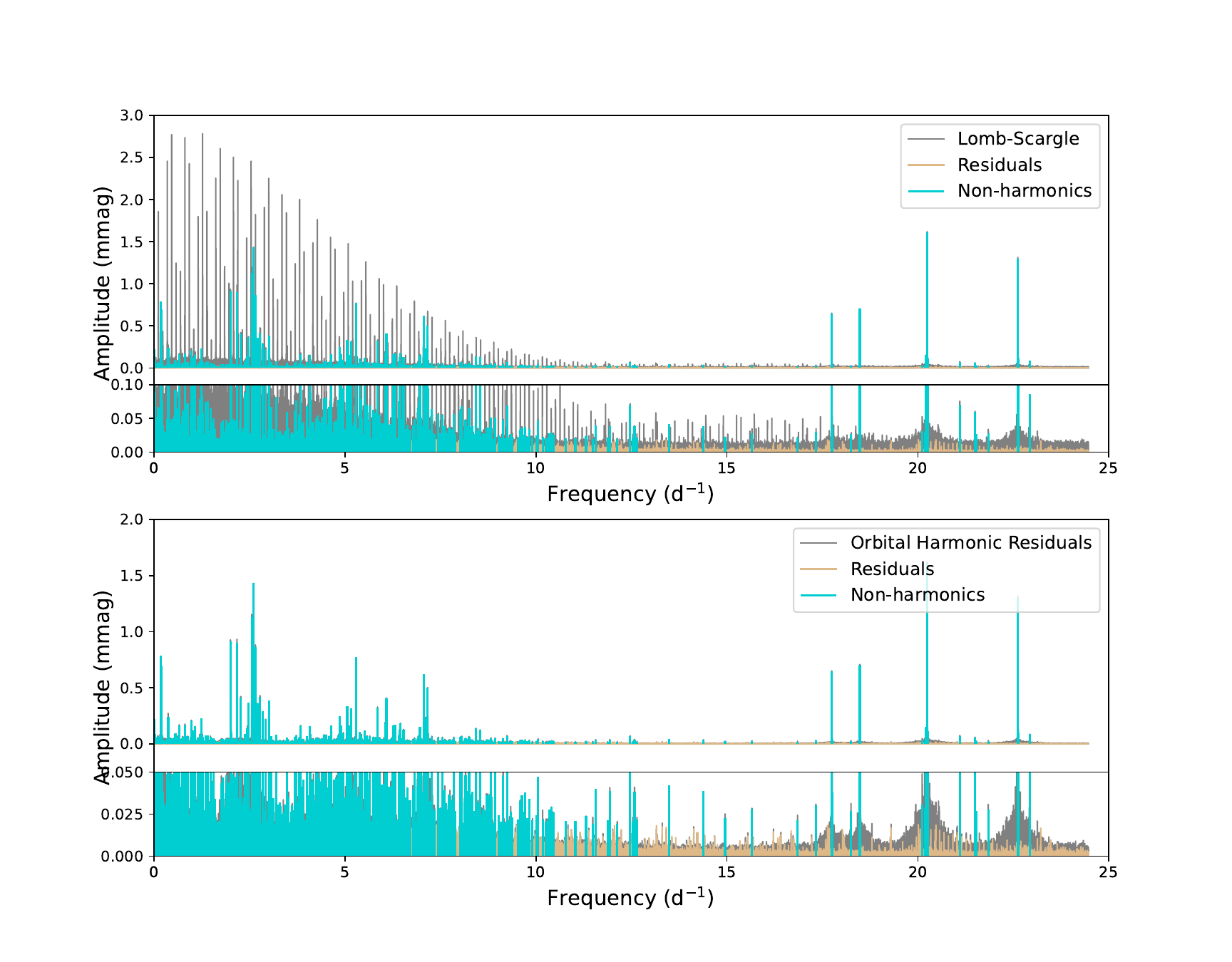}
       \caption{Lomb-Scargle periodogram of the normalised, detrended light curve (grey) with non-orbital harmonic frequencies extracted by \starshadow\ (light blue) and the periodogram of the residual light curve (orange). The lower panel is the periodogram of the light curve when the orbital harmonics (see Fig. \ref{fig:phasefold_sts}) are removed.}
    \label{fig:scg_sts}
\end{figure*}

\begin{figure*}
\centering
   \includegraphics[width=1.99\columnwidth]{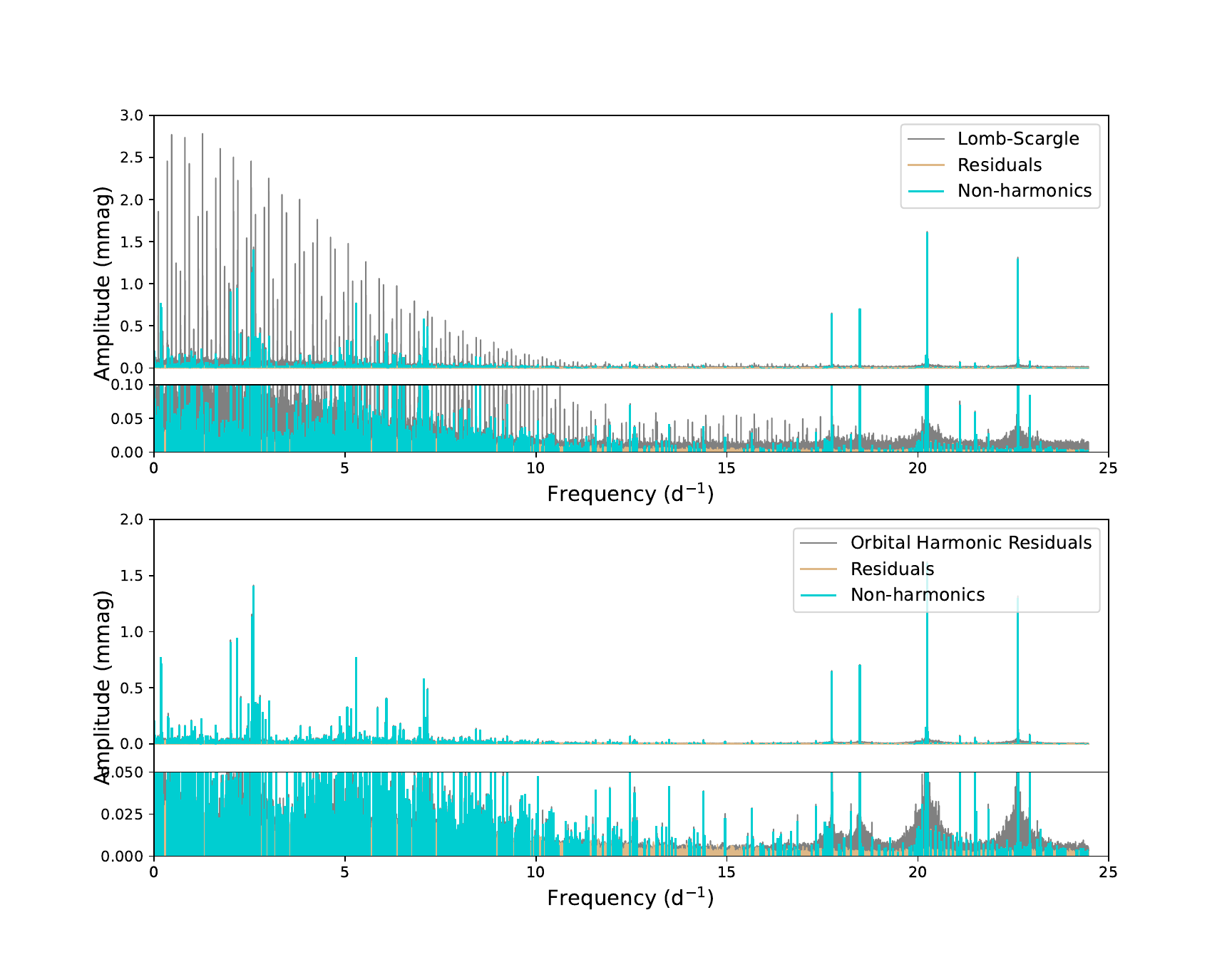}
       \caption{As Fig. \ref{fig:scg_sts}, but showing the \pofour\ frequency extraction.}
    \label{fig:scg_pofour}
\end{figure*}

\label{sec:lit_constraints}

In this section, we provide a brief summary of selected constraints on orbital and stellar properties from the literature relevant to the asteroseismic modelling of Aa. Much of the previous literature relating to KIC~4150611 is dedicated to constraining the orbital properties of the system. 
These constraints are essential to the identification and extraction of stellar oscillation frequencies as they allow confident identification of orbital harmonics (see Section \ref{sec:meth:extraction}). 

Using a variety of techniques, several works have provided estimates of the orbital periods of KIC~4150611's different components \citep{prsa2011,shibahashi2012,balona2014,rowe2015,heliminiak2017,kemp2024eclipse}. For the Ab, B, and C binaries, we make use of the orbital periods from \cite{heliminiak2017} to aid the identification of orbital harmonics.
The complex, time-variant geometry of the 94.2d eclipses precludes harmonic analysis; these eclipses are removed from the light curve in the time domain before the Fourier-space frequency analysis commences.

From \emph{Gaia} DR3 \citep{gaia2023}, we have an estimate for the distance to the system of $123.6^{+1.6}_{-2.3}$~pc\footnote{from Gaia's \protect \texttt{distance\_gspphot} label.}, which (using a bolometric\footnote{The bolometric correction is taken to be +0.01, see \protect \url{https://www.pas.rochester.edu/\%7Eemamajek/EEM_dwarf_UBVIJHK_colors_Teff.txt} \citep{pecaut2013}} magnitude of 8.03 for the system) in turn provides an estimate for the system's luminosity of $7.5 \pm 0.3$ L\solar\ (log($L$) = $0.875\pm0.0175$~dex). To make the step from a system luminosity to a stellar luminosity for Aa, we must consider the light fraction of Aa in the system.

The primary makes up most of the light in the system, although estimates of the exact light fraction vary. \cite{kemp2024eclipse} estimate a light fraction of between 0.92 and 0.94 based on spectroscopic analysis of TRES \citep{szentgyorgyi2007} spectra, but arrive at a lower value of roughly $0.84\pm0.03$ when considering only the eclipses, a value similar to the 0.85 light fraction obtained by \cite{heliminiak2017} in their eclipse modelling of the Ab, B, and C binaries. \cite{kemp2024eclipse} consider the effect of this uncertainty on their estimates of system's properties from their eclipse analysis, noting that the spectroscopic analysis appears far more sensitive to the light fraction than the eclipse modelling and therefore prefer a light fraction of around 0.92. A light fraction of 0.85 implies a \logl\ for Aa of $0.804\pm0.017$~dex, while a light fraction of 0.92 implies a \logl\ for Aa of $0.838\pm0.017$~dex. Note that the quoted uncertainties only propagate the uncertainty in parallax, and so are lower limits.

From the eclipse modelling in \cite{kemp2024eclipse}, the stellar radius for Aa was conservatively estimated as $1.64\pm0.06$~R\solar\ when considering the possibility of either a low or high light fraction being true. Each of \cite{kemp2024eclipse}'s two Markov Chain Monte Carlo (MCMC) simulations with constrained light fractions arrive at 1-$\sigma$ uncertainties of approximately 0.01~R\solar.

\cite{heliminiak2017} provide estimates of several of the properties of Aa properties using isochrone fits to Ba and Bb. These properties include the mass ($1.64\pm0.06$~M\solar), radius ($1.376\pm0.013$~R\solar), \logg\ ($4.38\pm0.01$) dex, and effective temperature ($8440\pm280$~K). However, isochrone fits suffer from high levels of modelling uncertainty, being tied to the underlying grid of stellar models. \cite{kemp2024eclipse} found poor agreement between their eclipse modelling and the stellar properties estimated from the isochrone fits of \cite{heliminiak2017}, including the radius and mass ratio estimates for the Aa, Ab1, and Ab2 components of the A triple. The effective temperature (\teff) of Aa from the isochrone fits is also too high compared to both the atmospheric modelling conducted in \cite{kemp2024eclipse} and previous spectroscopic analysis from \cite{niemczura2015}. The properties from the isochrone fits we use only for comparative purposes.

The atmospheric modelling on the disentangled spectra of Aa presented in \cite{kemp2024eclipse} estimate $T_{\rm eff} = 7280\pm70$ K and \logg\ = $4.14\pm0.18$ dex. From the eclipse modelling, there is a deviation from edge-on of at most 0.4$^{\circ}$, so the estimate of $v \sin i$ = $127\pm4$~\kms\ can be considered simply as the surface rotation. Combined with the conservative estimate of the stellar radius of $1.64\pm0.06$~R\solar\ from the eclipse modelling \citep{kemp2024eclipse}, this corresponds to an estimate on the rotation frequency of the surface, \rotfreqomegasurf, of $1.54\pm0.1$~\perd, where the quoted error encompasses the highest level of disagreement permissible within the 1-$\sigma$ of the radius and surface rotation constraints.

\section{Methodology}
\label{sec:meth}

\subsection{Frequency extraction}
\label{sec:meth:extraction}

The process of detrending and obtaining frequencies from KIC~4150611's complicated photometric time series was discussed in \cite{kemp2024eclipse}, but we summarise it again here, as it gives important context to three of the four period-spacing patterns we consider (see Section \ref{sec:meth:modelling}).

For the purpose of obtaining a g-mode period-spacing pattern for a \gdor\ star, \kepler's long-cadence time-series data can be considered ideal. KIC~4150611 is sparsely observed by TESS and has considerably poorer signal-to-noise, while short-cadence \kepler\ data offers little benefit to the low-frequency g-modes of \gdor\ stars but adds to the computational cost considerably.

Starting from the \kepler\ long-cadence target-pixel-files (TPFs), we employ a custom reduction and instrumental detrending following \cite{vanreeth2022,vanreeth2023}. The procedure is designed to improve the available signal while minimising contaminants from other stellar sources and avoid introducing signal from the instrumental detrending. This is achieved by applying a simple linear detrending model to each sector after a rough sinusoid model for the dominant physical effects (typically either eclipses or oscillations) has been subtracted. This is followed by outlier removal through a 5-$\sigma$ clipping and manual inspection. 
Throughout the detrending process, the 94.2d eclipses of the A triple were removed from the light-curve. In principle they could be reintroduced and detrended once the detrending curve was obtained, however, prewhitening in the frequency domain with these complicated eclipses included in the light curve is impractical and counterproductive. 

To obtain lists of frequencies, amplitudes, and phases we iteratively prewhiten the detrended light curve (with the 94.2d eclipses removed) using two different methods: manually using \pofour\footnote{\hyperlink{http://period04.net/}{http://period04.net/}}, and in an automated way using \starshadow\footnote{\url{https://github.com/LucIJspeert/star_shadow}} \citep{ijspeert2024}. Further details of this procedure can be found in Section 2.1.1 of \cite{kemp2024eclipse}, and the detrended light-curve -- and model fit using the \starshadow\ frequency model -- can be seen in Fig. \ref{fig:lightcurve_sts}. However, it is relevant to highlight certain key differences between the two processes.

The first is that while both methods identify orbital harmonics of the 8.65d, 1.52d, and 1.43d eclipses, only \starshadow\ couples the orbital frequencies for all identified harmonics, leading to a more precise extraction of orbital harmonics. This has relevance due to a near-perfect coincidence between an 8.65d orbital harmonic and one of the g-modes of KIC~41501611. Fig. \ref{fig:phasefold_sts} shows the phase-folded orbital periods using the \starshadow\ frequencies, and can be compared directly to Figure 3 from \cite{kemp2024eclipse} and Figure 2 from \cite{heliminiak2017}, highlighting the consistency of the eclipse extraction.

The second difference worth highlighting is that, as an automated method, \starshadow\ employs a strict amplitude-hinting procedure (see, for example, \citealt{vanbeeck2021}). This means that it proceeds with iterative frequency extraction in descending amplitude order, stopping as soon as the Bayes information criterion reduces by less than two when extracting the next frequency. In contrast, when prewhitening using \pofour\ the selection of the prewhitened frequencies is at the discretion of the user, as is the point at which to stop the process. For discussion on different prewhitening procedures and the influences they can have on the resulting frequency lists, the reader may refer to \cite{degroote2009} and \cite{vanbeeck2021}.

The presence of low-frequency noise in the Fourier transform (commonly referred to as `red noise') can cause frequency extraction to halt before reaching low amplitude -- but high SNR -- frequencies in the high-frequency domain. This results in the extraction of a lower number of frequencies overall, with high-frequency orbital harmonics clearly appearing in the residuals (see Figs. \ref{fig:scg_sts} and \ref{fig:scg_pofour}). The lack of these frequencies appear to have little bearing on the quality of the orbital harmonic models, however, nor on the quality of the pre-whitening in the g-mode regime.

Manual extraction using \pofour\ results in 178 orbital harmonics for the 8.65d orbital period, 18 orbital harmonics for the 1.52d orbital period, 19 orbital harmonics for the 1.43d orbital period, and 1238 other frequencies. \starshadow's frequency extraction procedure results in 160 orbital harmonics for the 8.65d orbital period, 15 orbital harmonics for the 1.52d orbital period, 12 orbital harmonics for the 1.43d orbital period, and 884 other frequencies. These frequency lists form the basis of our subsequent asteroseismic analysis.

\subsection{period-spacing pattern identification}
\label{sec:meth:ps_id}

The objective of this work is to provide asteroseismic analysis of the g-mode period-spacing pattern of KIC~4150611. This pattern is determined in \cite{li2020binary} to be the prograde dipole pattern. 
The low-frequency gravito-inertial modes detected in our target are often called prograde dipole modes, deriving from the limited notation $(l,m)=(1,+1)$ in terms of spherical harmonics for reasons of simplicity; in reality this spherical harmonic component delivers only the
dominant contribution to the actual Hough eigenfunction \citep{hough1898}
of such modes. For this reason,  
\citet{lee1997} introduced the more general notation $(k,m)=(0,1)$ for such modes, where $k\equiv l-|m|$.

We consider four different period-spacing patterns to evaluate how different extraction processes and pattern identification decisions can affect our results. The first pattern considered is the \cite{li2020binary} pattern, which we do not modify in any way. This pattern was also constructed from long-cadence \kepler\ data, but with a different light curve extraction and detrending process. A relatively low SNR threshold of 3 was applied when considering frequency significance by the algorithm employed by \cite{li2020binary}. We will refer to this pattern as the \patli\ pattern.

To construct period-spacing patterns from our own various light curves and their prewhitened frequency lists, we employ an iterative procedure relying on manually identifying candidate pattern members in the region of the Lomb-Scargle periodogram \citep{lomb1976,scargle1982} relevant to \gdor\ stars and (re-)fitting a theoretical asymptotic period-spacing pattern produced by \amigo \footnote{\url{https://github.com/TVanReeth/amigo}}\ \citep{vanreeth2016,vanreeth2018}, which relies on 
the numerical approximations worked out by \citet{Townsend2020}.



\amigo\ computes theoretical period-spacing patterns assuming a rigidly rotating, chemically homogeneous star under the traditional approximation of rotation (TAR, \citealt{eckart1960,berthomieu1978,lee1989,lee1997,mathis2009,bouabid2013,vanreeth2016}). Under the TAR, the horizontal component of the rotation vector is neglected, rendering the oscillation equations separable in spherical coordinates. The effect of rotation on the radial and azimuthal displacements are neglected while in the latitudinal direction they can be computed by solving the Laplace tidal equation, for which solutions can be expressed in terms of Hough functions \citep{hough1898}. This achieves the accuracy required for the modelling of observed low-frequency modes in rotating stars, in contrast to perturbative asymptotic predictions (see, for example, \citealt{shibahashi1979}). Indeed, given the significant frequency shifts induced by the Coriolis acceleration \citep[see, for example, Figures \,3 \& 4 in][]{aerts2023}, one should not  approximate gravito-inertial modes from computations in the perturbative regime.

The TAR is valid for modes where the displacement vector is dominant in the horizontal plane, which is an excellent approximation for $\gamma\,$Dor stars \citep{aerts2021}, including KIC~4150611. For the majority of these pulsators, the modes in the co-rotating frame have a frequency less than twice the rotation frequency, and the stellar rotation frequency is much less than both the Brunt-V\"{a}is\"{a}l\"{a} and Lamb frequencies, which govern the stability of buoyant and acoustic restoring forces, respectively, to allow for mode propagation \citep{aerts2023}.
This allows for asymptotic period-spacing patterns to be computed as a function of the rotation frequency in the near-core region \rotfreqomegacore\ and the buoyancy travel time $\Pi_0$ originally defined by \citet{tassoul1980}:

\begin{equation}
\Pi_0=2\pi^2 \Biggl(\int_{r_1}^{r_2} N \frac{\rm{d}r}{r} \Biggr) ^{-1},
\end{equation}
\noindent
where $N$ is the Brunt-V\"{a}is\"{a}l\"{a} frequency and $r_1$ and $r_2$ are the boundaries of the mode cavity. The frequencies of the g-modes belonging to a period-spacing pattern of known degree $l$ and azimuthal order $m$ can then be computed according to:

\begin{equation}
    f_{lmn} = \frac{\sqrt{\lambda_{lm,s}}}{(n+\alpha)\Pi_0} + m\Omega_{\rm c},
\end{equation}

where $\lambda_{lm,s}$ is the mode-specific eigenvalue of the Laplace tidal equation and $\alpha$ is a phase term determined by the details of the mode-cavity boundaries, usually treated as a free parameter (around 0.5 in \gdor\ stars) in practice \citep{bouabid2013,vanreeth2015method,vanreeth2016}. The value of $\lambda_{lm,s}$ is determined for each oscillation frequency by its spin parameter $s=2\Omega_{\rm c}/\omega_{lmn}$, where $\omega_{lmn}$ is the mode's angular frequency in the co-rotating reference frame.

\amigo\ determines $\lambda_{lm,s}$ using \texttt{GYRE}'s `\texttt{lambda(nu)}' function, first sampling a grid over a frequency domain appropriate to the star and then interpolating to determine precise  values of $\lambda_{lm,s}$ of each mode. In this way, \amigo\ is able to quickly produce asymptotic period-spacing patterns for high-order gravito-inertial modes as a function of \rotfreqomegacore\ and $\Pi_0$.

By computing a grid of such patterns varying \rotfreqomegacore\ and $\Pi_0$ and comparing with an observed period-spacing pattern, best-fitting values and uncertainty estimates for \rotfreqomegacore\ and $\Pi_0$ can be obtained without computing detailed stellar and asteroseismic models. In addition to being faster, this has the advantage of avoiding unpredictable systematic errors due to uncertainty in the details of the input physics of stellar models. However, by the same token, the links between these predictions  and the physics of stellar interiors is absent, limiting the inferences that can be drawn directly. Further, as a natural consequence of the underlying physical assumptions in the approach, structural information present in the pattern due to chemical gradients and other neglected physics cannot be reproduced. In stars where structural variations throughout the interior is large and induces deviations from a smooth period spacing pattern, the accuracy of the stellar parameters inferred of methods like \amigo\ will be limited. This is not a concern in the case of KIC~4150611, where we observe a very smooth period-spacing pattern.

In addition to its utility in providing estimates of \rotfreqomegacore and $\Pi_0$ (see Section \ref{sec:results_paraest}), pattern predictions from theoretical frameworks such as \amigo\ provide a useful aid in the identification of oscillation frequencies that may form part of a period-spacing pattern, as highlighted by \citet{vanreeth2015method}. Starting from obvious pattern-members, we use \amigo's pattern predictions as a guide to assist in identifying additional candidate pattern oscillations, which we then feed back into \amigo\ in an iterative process.



\begin{figure*}
\centering
   \includegraphics[width=1.9\columnwidth]{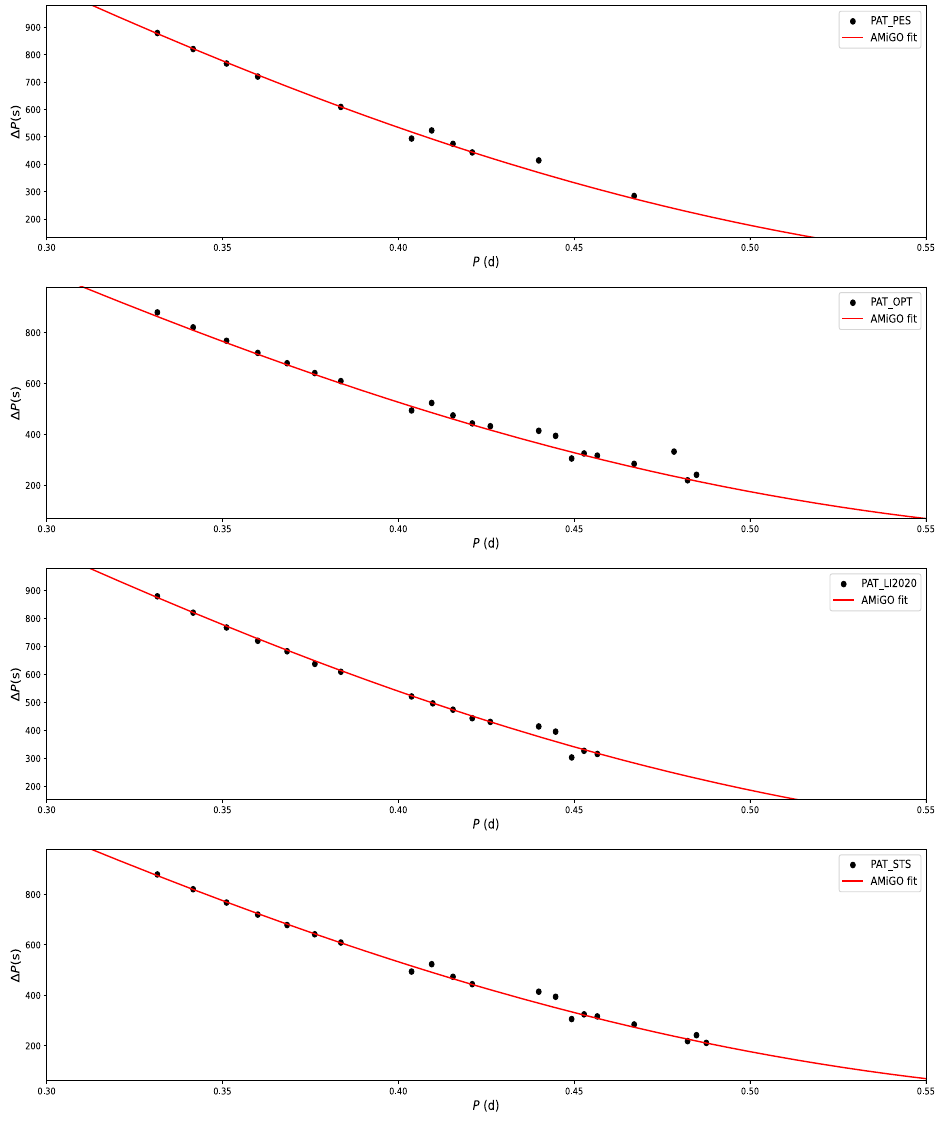}
       \caption{period-spacing patterns considered in this work, along with their \protect \amigo\ fits.}
    \label{fig:patshared}
\end{figure*}

In this manner, we arrive at three period-spacing patterns (in addition to the previously identified pattern in \cite{li2020binary}: two from the frequency list extracted using \pofour\ and one from the frequency list extracted using \starshadow. All patterns are included for ease of comparison in Fig. \ref{fig:patshared}.
The first pattern we obtain from the \pofour\ prewhitening process is extracted in a deliberately pessimistic manner, only taking frequencies that have SNR strictly above 5.6 \citep{baran2015} and are not ambiguous. The result is a pattern with several gaps and only short segments of consecutive modes, particularly at longer periods (equivalent to higher radial order $n$). This pattern represents a pessimistic pattern selection, and we will refer to it as \patpes.

The second pattern from the \pofour\ frequency list is obtained by selecting frequencies in a deliberately optimistic manner. In this selection, we allow moderate deviations below a SNR of 5.6, attempting to build the largest pattern possible. This includes judging whether a given mode is real and of stellar origin not only based on the mode's own properties -- such as its SNR -- but also making use of prior knowledge of the existing pattern. The result is a more complete period-spacing pattern that extends to higher radial order modes than both the \patli\ and \patpes\ patterns. We refer to this pattern as \patopt.

The final pattern makes use of the \starshadow\ frequency list and is also obtained in an optimistic manner, allowing moderate deviations in SNR for modes which fit the rest of the pattern. The resulting pattern is similar in length and completeness to \patopt, although the different extraction procedure results in differences to the extraction of low amplitude, high-order modes relevant to the pattern. We refer to this pattern as \patsts.

Further details about the extraction characteristics in the period region surrounding \patli, \patopt, \patpes\ and \patsts\ are shown visually in Figs. \ref{fig:patli2020}-\ref{fig:patsts}. The top panel of each figure shows the extracted frequencies, Lomb-Scargle periodogram for the relevant light curve with eclipse harmonics excluded, and the residual periodogram. The predicted frequencies from the \amigo\ fit to the pattern are also shown, as well as the SNR and estimated errors in amplitude and period for each extracted mode. In Fig. \ref{fig:patli2020}, the extracted frequencies, Lomb-Scargle periodogram, and residuals shown are from the \pofour\ frequency list for comparison with the \patli\ period-spacing pattern, which is plotted along the x-axis. In all figures, the lower panels show the period-spacing pattern and relevant \amigo\ fit to each pattern.

Note that while they are not shown in Figs. \ref{fig:patpes}-\ref{fig:patsts}, orbital harmonics, alias frequency predictions, and linear combinations of high-amplitude mode pairs were taken into account during pattern identification. Although there are several alias frequencies that fall within the domain of the period-spacing pattern, particularly those relating to the 94.2d orbit, none fall close enough to selected modes to be of concern.

There is a near-perfect coincidence between one of the 8.65d orbital harmonics and the extracted high-amplitude mode at 0.37d (this is excluded for this reason in the \patpes\ pattern, but included in the \patli, \patopt, and \patsts\ patterns). This is the cause of the nonphysical oscillatory behaviour that appears in the 8.65d phase-fold when using the \pofour\ frequency list to construct harmonic models of the eclipses discussed in \cite{kemp2024eclipse}. This is avoided in the \starshadow\ extraction due to the enforcement of frequency-coupling between orbital harmonics. However, when comparing these patterns in Figure \ref{fig:patshared} (see also Fig. \ref{fig:patopt} and \ref{fig:patsts}), it is clear that this makes little difference to the period-spacing pattern; the two patterns are essentially identical in the high-amplitude region below 0.4d.

All four patterns exhibit similar behaviour in the high-amplitude, short-period region. Differences only start to become noticeable when considering the low-amplitude, long-period modes. The \patli\ period-spacing pattern's second consecutive mode sequence (between approximately 0.4 and 0.43d) is very regular, but relies on extremely low-amplitude modes. Interestingly, it is the location of the second mode in this sequence -- one of the relatively high SNR modes -- that is the most noteworthy deviation when comparing with the \patopt\ and \patsts\ patterns. This mode is shifted slightly towards a lower period in both the \starshadow\ and \pofour\ frequency lists. The structure of the third consecutive sequence, between 0.43d and 0.46d, is essentially identical between all frequency lists.

\patpes, \patopt, and \patsts\ all include modes beyond the third and final sequence of \patli, although in the case of \patpes\ this includes very few consecutive modes. Beyond the high amplitude peak at approximately 0.46d, there is a low amplitude valley followed by a high-amplitude peak. Several of these modes do not satisfy SNR$>5.6$, and the predicted mode density in this region is approaching levels where chance coincidences of individual observed periods with the predicted would be a concern. Nonetheless, a series of modes with spacings consistent with the \amigo\ predictions can be found here, noting that extracted frequencies in this region are different in places between the \patsts\ pattern and the \patopt\ pattern to the point where it appreciably affects the structure.

At this point, it is worth noting how the \amigo\ fit predictions are used. Simply comparing the period predictions to the locations of extracted modes is often unhelpful, as individual oscillation periods can be affected by the chemical gradients inside the star. 
These period shifts manifest as structural glitches in the period-spacing pattern before quickly returning to the asymptotic behaviour, and are not able to be modelled by \amigo. The structural glitch seen in the third consecutive sequence in \patli, \patopt, and \patsts\ is an example of such a mode shift (around radial order 34 in Fig. \ref{fig:bfcthree}).



It is clear that the inclusion -- or not -- of these high-order modes in the pattern has a much larger effect on the \amigo\ fits than the differences in the actual mode frequencies between the different patterns. \patli\ and \patpes\ have relatively similar \amigo\ fits, which rely far more upon the highest amplitude, short-period modes. Conversely, despite the structural differences between \patopt\ and \patsts\ at high radial orders, the fit between the two is near identical. As \amigo\ only includes the physics needed for modelling the overall pattern shape, this is anticipated.


\subsection{Asteroseismic grid modelling}
\label{sec:meth:modelling}

\begin{table}[]

\caption{External constraints and labelling conventions}
\begin{tabular}{l|lll}
Label & \teff\ (K) & \logg\ (dex) & $R$ (R\solar) \\ \hline
All & - & - & - \\
Spectro & 7130-7430 & 3.84-4.44 & - \\
R & - & - & 1.55-1.75 \\
R and Spectro & 7130-7430 & 3.84-4.44 & 1.55-1.75 \\
Tight R & - & - & 1.63-1.65 \\
Tight R and Spectro & 7130-7430 & 3.84-4.44 & -
\end{tabular}
\label{tab:constraints}
\tablefoot{The default external constraints are formed by doubling the error estimates from \cite{kemp2024eclipse}. The `Tight R' constraint is closer to the formal 1-$\sigma$ uncertainty under a spectroscopically constrained light fraction (see \citealt{kemp2024eclipse} for details).}
\end{table}

The most expensive aspect of asteroseismic modelling using gravity modes is undoubtedly the computation of grids of stellar models and their pulsations. In order for the resolution of the grid not to adversely affect the results, dense stellar grids are required. These grids must, at minimum, span the zero-age main sequence (ZAMS) mass and an age proxy such as the central hydrogen mass fraction, \xc. As the tilt of a period-spacing pattern is highly sensitive to the near-core rotation frequency \rotfreqomegacore, dense sampling of this parameter is required when computing the pulsations if the pattern is to be compared with directly. The required sampling density of \rotfreqomegacore\ can be reduced by optimising the rotation scaling independently as part of the merit function, essentially finding the most probable \rotfreqomegacore\ for each structural model (see \citealt{michielsen2023}). However, probing other additional stellar physics such as convective boundary mixing parameters (e.g., the degree of step \citep{shaviv1973,zahn1991} or exponential \citep{freytag1996,herwig2000} overshooting) each involve adding another dimension to a grid of stellar models that multiplies the computational cost.

In this work, we make use of \cthree\footnote{\url{https://github.com/JMombarg/c3po}}, a neural network-driven asteroseismic modelling code for \gdor\ stars \citep{mombarg2021}. It rapidly makes  predictions for the pulsation frequencies, period-spacings, and radial-orders of a star based on its mass \M, \xc, metallicity \Z, \rotfreqomegacore, and degree of exponential core overshooting \fov. This is accomplished by taking the average frequency prediction from five different neural networks dedicated to this task. A separate neural network computes \teff\ and \logg, common physical constraints from spectroscopy, while another computes the luminosity \lum\ for each model, which can then be compared with astrometric luminosities such as those from \emph{Gaia} \citep{gaia2016}. The underlying stellar structure and pulsation models used to train \cthree\ were computed using \texttt{MESA} \citep{paxton2013} and \texttt{GYRE} \citep{townsend2013} respectively. 
Despite not including rotation explicitly, the training set used for \cthree\ used stellar models computed with a constant envelope mixing ($D_\mathrm{mix}=1 $\,cm$^2$\,s$^{-1}$) to account for transport processes in the envelope. This low level of $D_\mathrm{mix}$ reflects the low levels of envelope mixing found in \gdor\ stars \citep{vanreeth2016,mombarg2019}. The training set includes a diffusive-exponential core overshoot prescription, varying the diffusive exponential overshoot factor $f_\mathrm{ov}$ between 0.01-0.03. The training set spans the \gdor\ range in mass (1.3 M\solar -- 2.0 M\solar) and spans a small range of metallicities around solar (\Z=0.011-0.015). 
Further details of the underlying physics of the \texttt{MESA} models and \texttt{GYRE} pulsation models on which the network is trained can be found in \cite{mombarg2021}.

In addition to the neural network predictor modules, \cthree\ includes a modelling framework that handles both radial-order matching using the neural-network's pulsation predictions and assigning merit values for each model. The period-spacing patterns are built by matching the first period in the longest sequence of consecutive modes, and building out the pattern to the other sequences from there. The merit function includes both the periods and the period spacings; this can be viewed as a compromise between constraining power and sensitivity to un-modelled physics.
\cite{michielsen2021} find that, for their case-study star of KIC~7760680, a slowly pulsating B (SPB) star with an exceptionally long period-spacing pattern with a high level of structure, a merit function accounting only for the periods is inferior to one accounting only for mode period spacings, citing high variance in theoretical mode predictions for the pulsation periods.
By combining both period-spacings and periods, \cthree's methodology attempts to ensure that period-spacings are accurately fit, while rewarding models that also have agreement in the mode periods.

\cite{michielsen2021} also compare two merit functions: the commonly used \chisqr\ and the Mahalanobis distance (MD, \citealt{johnson2019}) with an additional term for the model variance \citep{aerts2018}. This additional term adjusts the weight given to each observation according to both the degree of covariance between each observable according to the model grid -- in this way attempting to measure the modelling uncertainty -- on top of the covariance between the observables themselves. This results in a broadening of the parameter-space, widening uncertainty regions. \cite{michielsen2021} find that the MD-based merit function outperforms the \chisqr\ merit function insofar as it arrives at more realistic uncertainties. However, an important caveat is that KIC~7760680 had an unusually long and complete period-spacing pattern with prominent structures that could be mostly explained by physics included within the underlying pulsation models.

Preliminary analysis on KIC~4150611 making use of the forward modelling software package \texttt{Foam} \citep{michielsen2024foam} revealed that use of the MD-based merit function led to poor fits to the period-spacing pattern regardless of whether the period-spacing or the periods were fit. Patterns which vaguely approximated the structural glitches in the period-spacing pattern strongly preferred despite those models utterly failing to reproduce the rest of the pattern. The \chisqr\ statistic, on the other hand, placed no special weight on those structures, instead favouring patterns which matched the observed pattern as a whole.

\cthree\ supports both the \chisqr\ and the MD-based merit function merit functions; we make use of the \chisqr\ for KIC~4150611. \cthree's modelling framework also supports using external constraints such as \teff, \logg, and \lum\ in the sampling phase, which can allow for more efficient sampling of the relevant space. We do not make use of this feature, instead opting to compute many models uniformly distributed across the entire \gdor\ range. This is to facilitate discussion of the relative constraining power of asteroseismic, spectroscopic, and eclipse-modelling observables and their combinations for different stellar parameters. To assess the impact of sampling effects on our results, we compute our results using two different samples for each pattern: a medium resolution sample, with \timestento{8}{5} models across the \gdor\ range, and a high resolution sample, with \timestento{3}{6} models. We present the results from the high resolution sampling in the main text, while the results from the alternative `Tight R' radius constraint in the appendices. All figures relating to the medium resolution sampling are included in the online supplementary material. A summary of the different configurations of external constraints considered is presented in Table \ref{tab:constraints}.

We provide an estimate of the uncertainty for our \cthree\ modelling by estimating a $1-\sigma$ $\chi^2$ cut-off using equation 9 from \cite{mombarg2021}, and then taking the maximum and minimum parameter values from that distribution to form the $1-\sigma$ estimate. These uncertainty estimates should be treated with caution, particularly for cases where the parameters do not have normal (or even quasi-normal) $\chi^2$ envelopes. Further, the $1-\sigma$ sub-samples these margins are based on contain relatively few (20-40) models in the most constraining case, adding an element of stochasticity to their estimation.

\subsection{MCMC parameter-based grid-search.}
\label{sec:meth:mcmc}
To better understand the correlations between and the uncertainty structure of our parameter estimates, we also employed a classical MCMC grid search. This was done using several different grids of stellar models (grids A-E): 

\begin{itemize}
\item Grid A: a grid composed of the lowest metallicity non-rotating models from \cite{mombarg2021} (the same underlying models used to train \cthree).
\item Grid B: a grid composed of solar-metallicity, rotating stellar models from \cite{mombarg2024} (employing hydrodynamic envelope mixing and angular momentum transport based on the rotation, as well as microscopic diffusion).
\item Grid C: a grid of rotating solar metallicity models from \cite{mombarg2024gaia} (employing rotational mixing but with fixed viscosity and no microscopic diffusion).
\item Grid D: a grid of \Z=0.0045 models also from \cite{mombarg2024gaia}.
\item Grid E: a grid of \Z=0.008 models with physics equivalent to \cite{mombarg2024gaia} \citep{aerts2024}.
\end{itemize}

\noindent Note that although the grid of low-metallicity models from \cite{mombarg2021} is from the same set used to train \cthree, it is considerably denser, as \cthree\ was trained only on the sub-set of stellar models where \texttt{GYRE} pulsation models were computed.

The grid search employed an MCMC sampling method similar to \cite{fritzewski2024} based on the external constraints $T_\mathrm{eff}$, $R$, $L$, $\log g$, and the asteroseismic measurement of the buoyancy travel time parameter $\Pi_0$. For the rotating stellar models, the asteroseismic near-core rotation \rotfreqomegacore\ was also used. We allowed the MCMC walkers to explore the grid in three unconstrained parameters $M$, $f_\mathrm{ov}$, and $X_c$ while keeping the envelope mixing fixed to $D_\mathrm{mix}=1$\,cm$^2$\,s$^{-1}$. We assumed a flat prior for all parameters.

\section{Results}
\label{sec:results}

In this section, we present the results of our asteroseismic modelling of the g-mode period-spacing pattern of KIC~4150611 Aa.

\begin{figure*}
\centering
\begin{subfigure}{0.99\columnwidth}
\centering
\includegraphics[width=0.99\columnwidth]{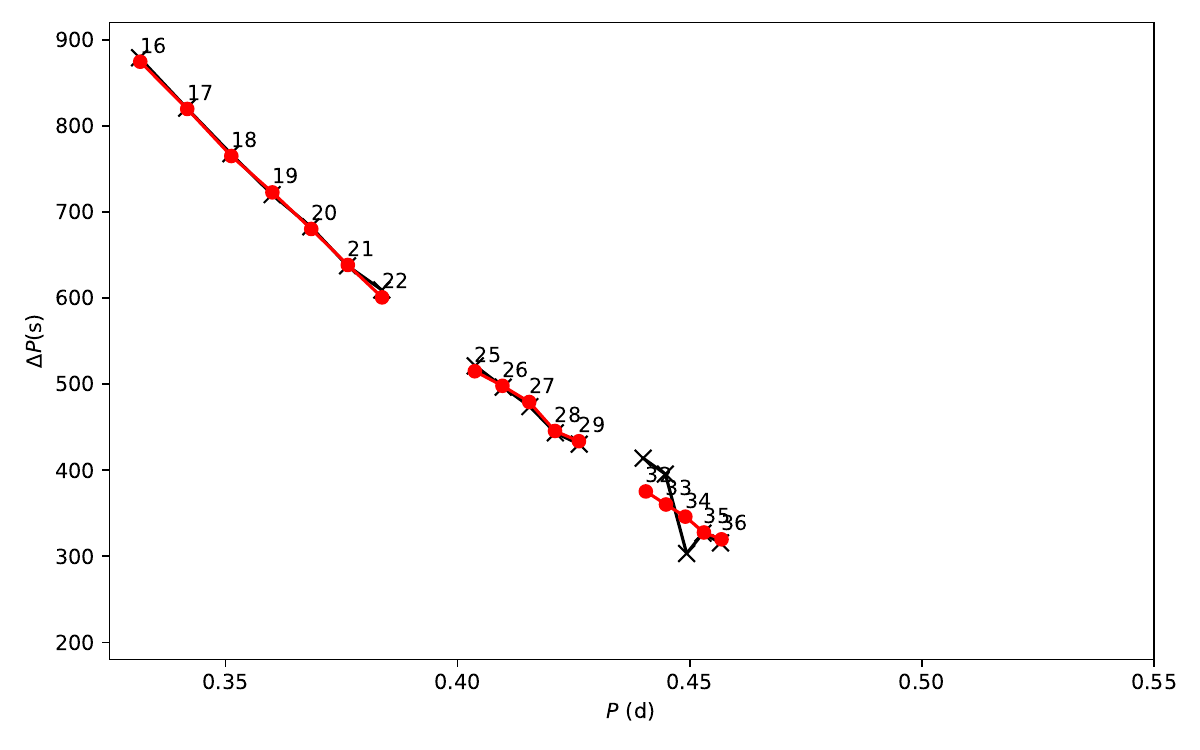}
\caption{\patli}
\label{fig:bfcthreepatli}
\end{subfigure}%
\begin{subfigure}{0.99\columnwidth}
\centering
\includegraphics[width=0.99\columnwidth]{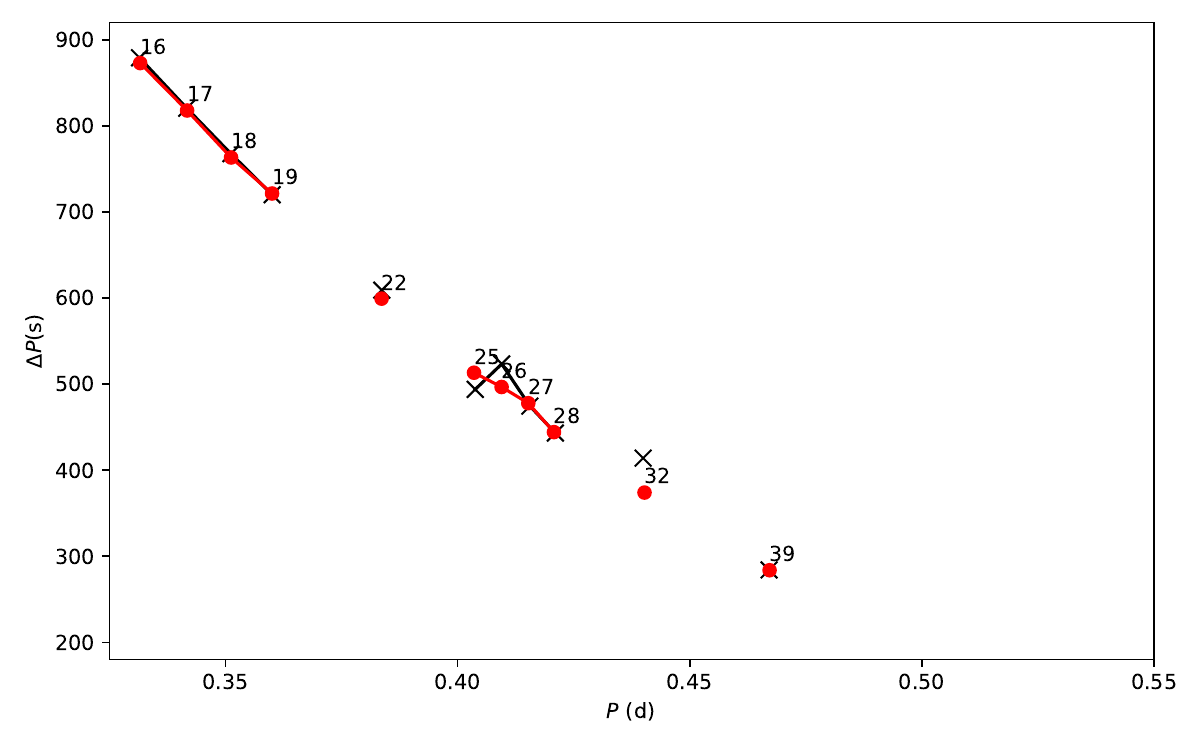}
\caption{\patpes}
\label{fig:bfcthreepatpes}
\end{subfigure}

\begin{subfigure}{0.99\columnwidth}
\centering
\includegraphics[width=0.99\columnwidth]{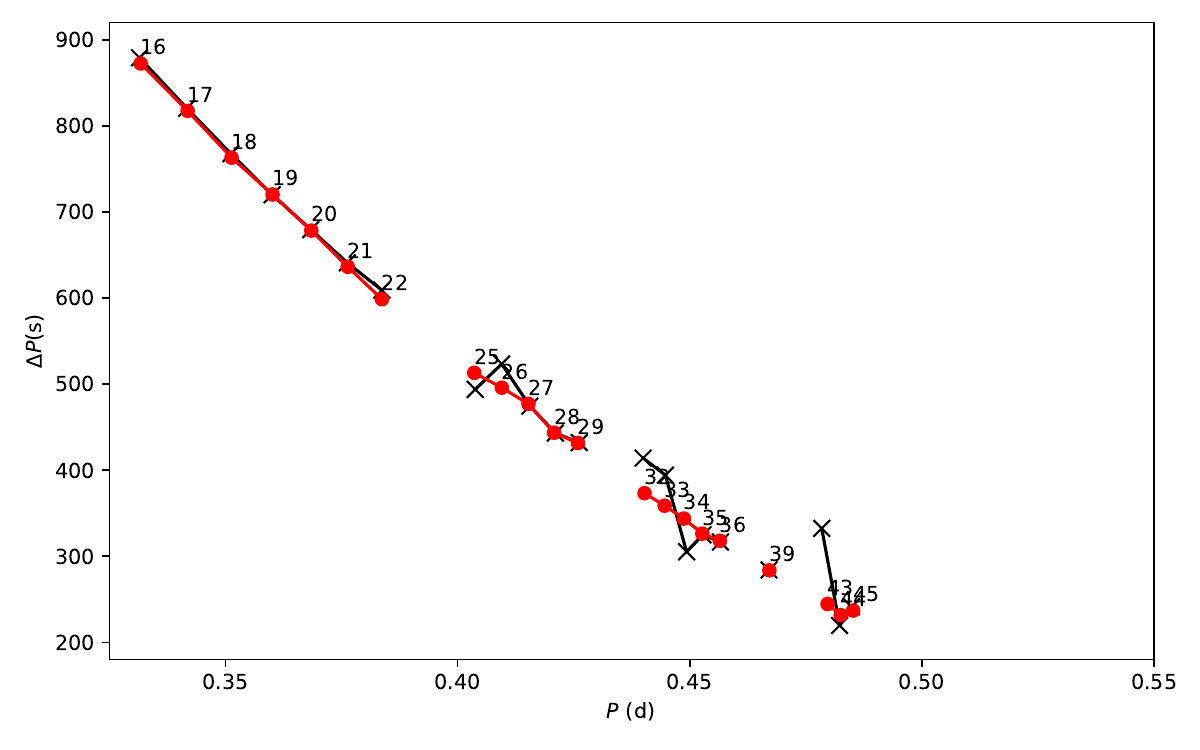}
\caption{\patopt}
\label{fig:bfcthreepatopt}
\end{subfigure}%
\begin{subfigure}{0.99\columnwidth}
\centering
\includegraphics[width=0.99\columnwidth]{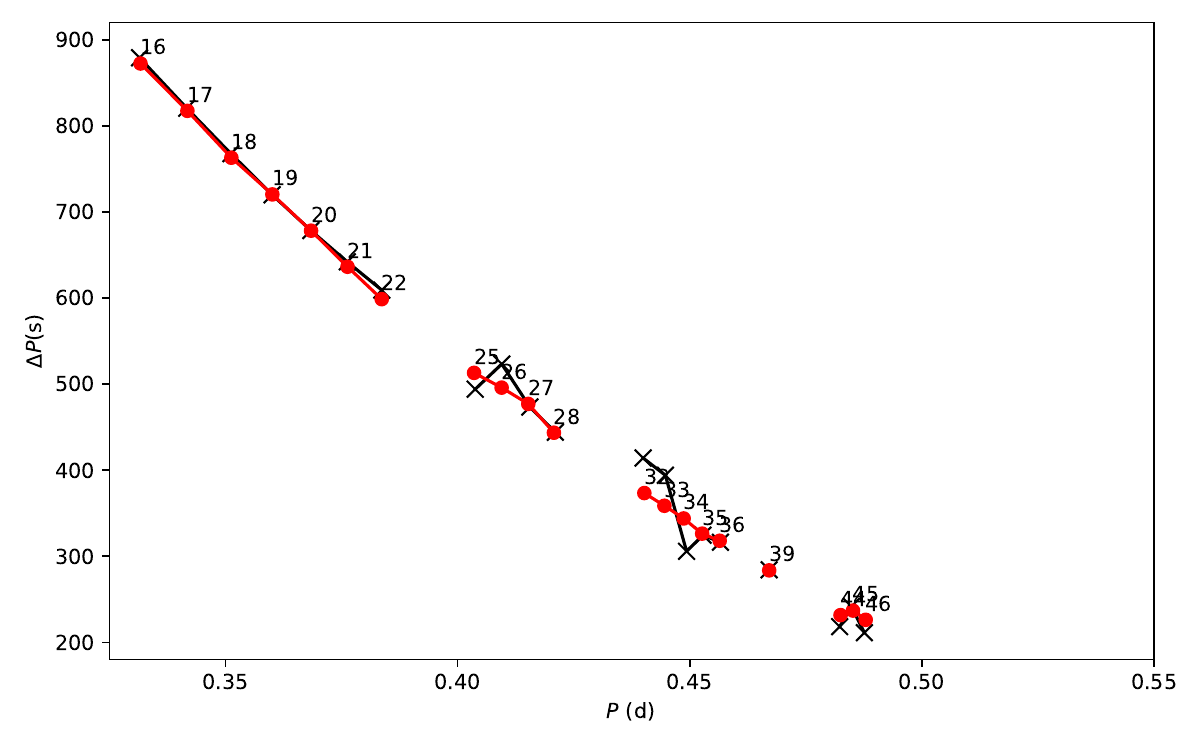}
\caption{\patsts}
\label{fig:bfcthreepatsts}
\end{subfigure}
\caption{Best-fitting \cthree\ models (red) consistent with the radial and spectroscopic constraints (see Table \ref{tab:bestcthree_highres}). Observed patterns are shown in black.}
\label{fig:bfcthree}
\end{figure*}

\begin{table*}[]
\caption{Parameters of the best-fitting models from \cthree's high resolution sample for each pattern under different external constraints.}
\resizebox{\textwidth}{!}{%
\bgroup
\def\arraystretch{1.5}
\begin{tabular}{ll|llllllll}
Pattern & Sample & \M & \xc & $R$ & \fov & \Z & log(\teff) & \logg & log(\lum) \\
\hline
 &  & M\solar & - & R\solar & - & - & dex & dex & dex \\
\hline
\patli & All & $1.958^{+0.02}_{-0.622}$ & $0.07^{+0.625}_{-0.018}$ & $3.051^{+0.86}_{-1.793}$ & $0.011^{+0.004}_{-0.0}$ & $0.01^{+0.019}_{-0.0}$ & $3.8554^{+0.0}_{-0.062}$ & $3.7612^{+0.6032}_{-0.2299}$ & $1.343^{+0.0295}_{-0.8698}$ \\
\patli & Spectro & $1.958^{+0.02}_{-0.566}$ & $0.07^{+0.569}_{-0.017}$ & $3.051^{+0.164}_{-1.698}$ & $0.011^{+0.003}_{-0.0}$ & $0.01^{+0.005}_{-0.0}$ & $3.8554^{+0.0}_{-0.0109}$ & $3.7612^{+0.5581}_{-0.0432}$ & $1.343^{+0.0216}_{-0.7621}$ \\
\patli & R & $1.478^{+0.042}_{-0.1}$ & $0.471^{+0.013}_{-0.09}$ & $1.589^{+0.16}_{-0.038}$ & $0.011^{+0.004}_{-0.0}$ & $0.011^{+0.015}_{-0.0}$ & $3.8537^{+0.0009}_{-0.0241}$ & $4.2057^{+0.0113}_{-0.0979}$ & $0.7629^{+0.0752}_{-0.0782}$ \\
\patli & R and Spectro & $1.478^{+0.042}_{-0.035}$ & $0.471^{+0.013}_{-0.084}$ & $1.589^{+0.158}_{-0.038}$ & $0.011^{+0.003}_{-0.0}$ & $0.011^{+0.004}_{-0.0}$ & $3.8537^{+0.0009}_{-0.0088}$ & $4.2057^{+0.0113}_{-0.0788}$ & $0.7629^{+0.0752}_{-0.0391}$ \\
\patli & Tight R & $1.459^{+0.061}_{-0.081}$ & $0.424^{+0.06}_{-0.043}$ & $1.644^{+0.104}_{-0.093}$ & $0.014^{+0.001}_{-0.003}$ & $0.011^{+0.015}_{-0.0}$ & $3.8466^{+0.0079}_{-0.017}$ & $4.1703^{+0.0467}_{-0.0624}$ & $0.7656^{+0.0725}_{-0.0809}$ \\
\patli & Tight R and Spectro & $1.459^{+0.061}_{-0.016}$ & $0.424^{+0.06}_{-0.037}$ & $1.644^{+0.103}_{-0.093}$ & $0.014^{+0.0}_{-0.003}$ & $0.011^{+0.004}_{-0.0}$ & $3.8466^{+0.0079}_{-0.0018}$ & $4.1703^{+0.0467}_{-0.0434}$ & $0.7656^{+0.0725}_{-0.0418}$ \\
\hline
\patpes & All & $1.693^{+0.297}_{-0.367}$ & $0.185^{+0.512}_{-0.132}$ & $2.375^{+1.014}_{-1.099}$ & $0.013^{+0.002}_{-0.002}$ & $0.01^{+0.009}_{-0.0}$ & $3.8434^{+0.0112}_{-0.019}$ & $3.9158^{+0.4379}_{-0.25}$ & $1.0765^{+0.2985}_{-0.5963}$ \\
\patpes & Spectro & $1.471^{+0.519}_{-0.091}$ & $0.495^{+0.149}_{-0.436}$ & $1.553^{+1.612}_{-0.212}$ & $0.011^{+0.004}_{-0.0}$ & $0.01^{+0.005}_{-0.0}$ & $3.8544^{+0.0002}_{-0.0099}$ & $4.2234^{+0.1}_{-0.4869}$ & $0.7447^{+0.6304}_{-0.1775}$ \\
\patpes & R & $1.471^{+0.05}_{-0.04}$ & $0.495^{+0.0}_{-0.098}$ & $1.553^{+0.175}_{-0.003}$ & $0.011^{+0.004}_{-0.0}$ & $0.01^{+0.006}_{-0.0}$ & $3.8544^{+0.0002}_{-0.0126}$ & $4.2234^{+0.0}_{-0.0857}$ & $0.7447^{+0.0927}_{-0.0363}$ \\
\patpes & R and Spectro & $1.471^{+0.05}_{-0.032}$ & $0.495^{+0.0}_{-0.098}$ & $1.553^{+0.175}_{-0.003}$ & $0.011^{+0.004}_{-0.0}$ & $0.01^{+0.005}_{-0.0}$ & $3.8544^{+0.0002}_{-0.0084}$ & $4.2234^{+0.0}_{-0.0781}$ & $0.7447^{+0.0927}_{-0.0194}$ \\
\patpes & Tight R & $1.494^{+0.027}_{-0.063}$ & $0.437^{+0.058}_{-0.04}$ & $1.65^{+0.078}_{-0.1}$ & $0.011^{+0.004}_{-0.0}$ & $0.011^{+0.005}_{-0.001}$ & $3.8546^{+0.0}_{-0.0128}$ & $4.1778^{+0.0456}_{-0.04}$ & $0.7952^{+0.0421}_{-0.0869}$ \\
\patpes & Tight R and Spectro & $1.494^{+0.027}_{-0.055}$ & $0.437^{+0.058}_{-0.04}$ & $1.65^{+0.078}_{-0.1}$ & $0.011^{+0.004}_{-0.0}$ & $0.011^{+0.003}_{-0.001}$ & $3.8546^{+0.0}_{-0.0086}$ & $4.1778^{+0.0456}_{-0.0325}$ & $0.7952^{+0.0421}_{-0.07}$ \\
\hline
\patopt & All & $1.512^{+0.487}_{-0.187}$ & $0.424^{+0.275}_{-0.374}$ & $1.678^{+2.586}_{-0.43}$ & $0.011^{+0.004}_{-0.0}$ & $0.01^{+0.02}_{-0.0}$ & $3.8558^{+0.0086}_{-0.0639}$ & $4.1681^{+0.1995}_{-0.6915}$ & $0.8188^{+0.6574}_{-0.3604}$ \\
\patopt & Spectro & $1.512^{+0.477}_{-0.127}$ & $0.424^{+0.273}_{-0.368}$ & $1.678^{+1.689}_{-0.363}$ & $0.011^{+0.004}_{-0.0}$ & $0.01^{+0.02}_{-0.0}$ & $3.8558^{+0.0086}_{-0.0115}$ & $4.1681^{+0.1762}_{-0.488}$ & $0.8188^{+0.5865}_{-0.259}$ \\
\patopt & R & $1.512^{+0.047}_{-0.094}$ & $0.424^{+0.125}_{-0.053}$ & $1.678^{+0.069}_{-0.126}$ & $0.011^{+0.004}_{-0.0}$ & $0.01^{+0.02}_{-0.0}$ & $3.8558^{+0.0086}_{-0.0147}$ & $4.1681^{+0.0591}_{-0.0481}$ & $0.8188^{+0.059}_{-0.1058}$ \\
\patopt & R and Spectro & $1.512^{+0.047}_{-0.071}$ & $0.424^{+0.125}_{-0.035}$ & $1.678^{+0.069}_{-0.126}$ & $0.011^{+0.004}_{-0.0}$ & $0.01^{+0.02}_{-0.0}$ & $3.8558^{+0.0086}_{-0.011}$ & $4.1681^{+0.0591}_{-0.0405}$ & $0.8188^{+0.059}_{-0.0873}$ \\
\patopt & Tight R & $1.492^{+0.067}_{-0.074}$ & $0.441^{+0.108}_{-0.07}$ & $1.639^{+0.109}_{-0.086}$ & $0.012^{+0.003}_{-0.001}$ & $0.01^{+0.02}_{-0.0}$ & $3.8556^{+0.0089}_{-0.0145}$ & $4.1831^{+0.0441}_{-0.0631}$ & $0.7925^{+0.0853}_{-0.0795}$ \\
\patopt & Tight R and Spectro & $1.492^{+0.067}_{-0.051}$ & $0.441^{+0.108}_{-0.052}$ & $1.639^{+0.109}_{-0.086}$ & $0.012^{+0.003}_{-0.001}$ & $0.01^{+0.02}_{-0.0}$ & $3.8556^{+0.0089}_{-0.0108}$ & $4.1831^{+0.0441}_{-0.0555}$ & $0.7925^{+0.0853}_{-0.061}$ \\
\hline
\patsts & All & $1.506^{+0.491}_{-0.164}$ & $0.434^{+0.26}_{-0.376}$ & $1.659^{+1.544}_{-0.394}$ & $0.011^{+0.004}_{-0.0}$ & $0.01^{+0.005}_{-0.0}$ & $3.8561^{+0.0023}_{-0.0252}$ & $4.1764^{+0.1851}_{-0.4517}$ & $0.8087^{+0.5706}_{-0.3264}$ \\
\patsts & Spectro & $1.506^{+0.491}_{-0.129}$ & $0.434^{+0.235}_{-0.376}$ & $1.659^{+1.544}_{-0.342}$ & $0.011^{+0.004}_{-0.0}$ & $0.01^{+0.004}_{-0.0}$ & $3.8561^{+0.0023}_{-0.0118}$ & $4.1764^{+0.1617}_{-0.4503}$ & $0.8087^{+0.5706}_{-0.2635}$ \\
\patsts & R & $1.506^{+0.0}_{-0.075}$ & $0.434^{+0.034}_{-0.045}$ & $1.659^{+0.068}_{-0.09}$ & $0.011^{+0.004}_{-0.0}$ & $0.01^{+0.004}_{-0.0}$ & $3.8561^{+0.0}_{-0.0137}$ & $4.1764^{+0.03}_{-0.0443}$ & $0.8087^{+0.0062}_{-0.0825}$ \\
\textbf{\patsts} & \textbf{R and Spectro} & $\mathbf{1.506^{+0.0}_{-0.071}}$ & $\mathbf{0.434^{+0.034}_{-0.045}}$ & $\mathbf{1.659^{+0.068}_{-0.09}}$ & $\mathbf{0.011^{+0.004}_{-0.0}}$ & $\mathbf{0.01^{+0.004}_{-0.0}}$ & $\mathbf{3.8561^{+0.0}_{-0.0116}}$ & $\mathbf{4.1764^{+0.03}_{-0.0443}}$ & $\mathbf{0.8087^{+0.0062}_{-0.0807}}$ \\
\patsts & Tight R & $1.455^{+0.051}_{-0.024}$ & $0.434^{+0.034}_{-0.045}$ & $1.632^{+0.094}_{-0.063}$ & $0.014^{+0.001}_{-0.003}$ & $0.012^{+0.003}_{-0.001}$ & $3.8463^{+0.0098}_{-0.0039}$ & $4.1756^{+0.0308}_{-0.0434}$ & $0.7589^{+0.056}_{-0.0327}$ \\
\patsts & Tight R and Spectro & $1.455^{+0.051}_{-0.02}$ & $0.434^{+0.034}_{-0.045}$ & $1.632^{+0.094}_{-0.063}$ & $0.014^{+0.001}_{-0.003}$ & $0.012^{+0.003}_{-0.001}$ & $3.8463^{+0.0098}_{-0.0018}$ & $4.1756^{+0.0308}_{-0.0434}$ & $0.7589^{+0.056}_{-0.0309}$
\end{tabular}
\egroup}
\tablefoot{The externally constrained \patsts\ pattern is the preferred solution, and marked in bold. The units of \teff, $g$, and $L$ are K, g$\,$cm$^{-3}$, and L\solar. The sub-samples are described in Table \ref{tab:constraints}.}
\label{tab:bestcthree_highres}
\end{table*}

\begin{figure*}
\centering
   \includegraphics[width=1.9\columnwidth]{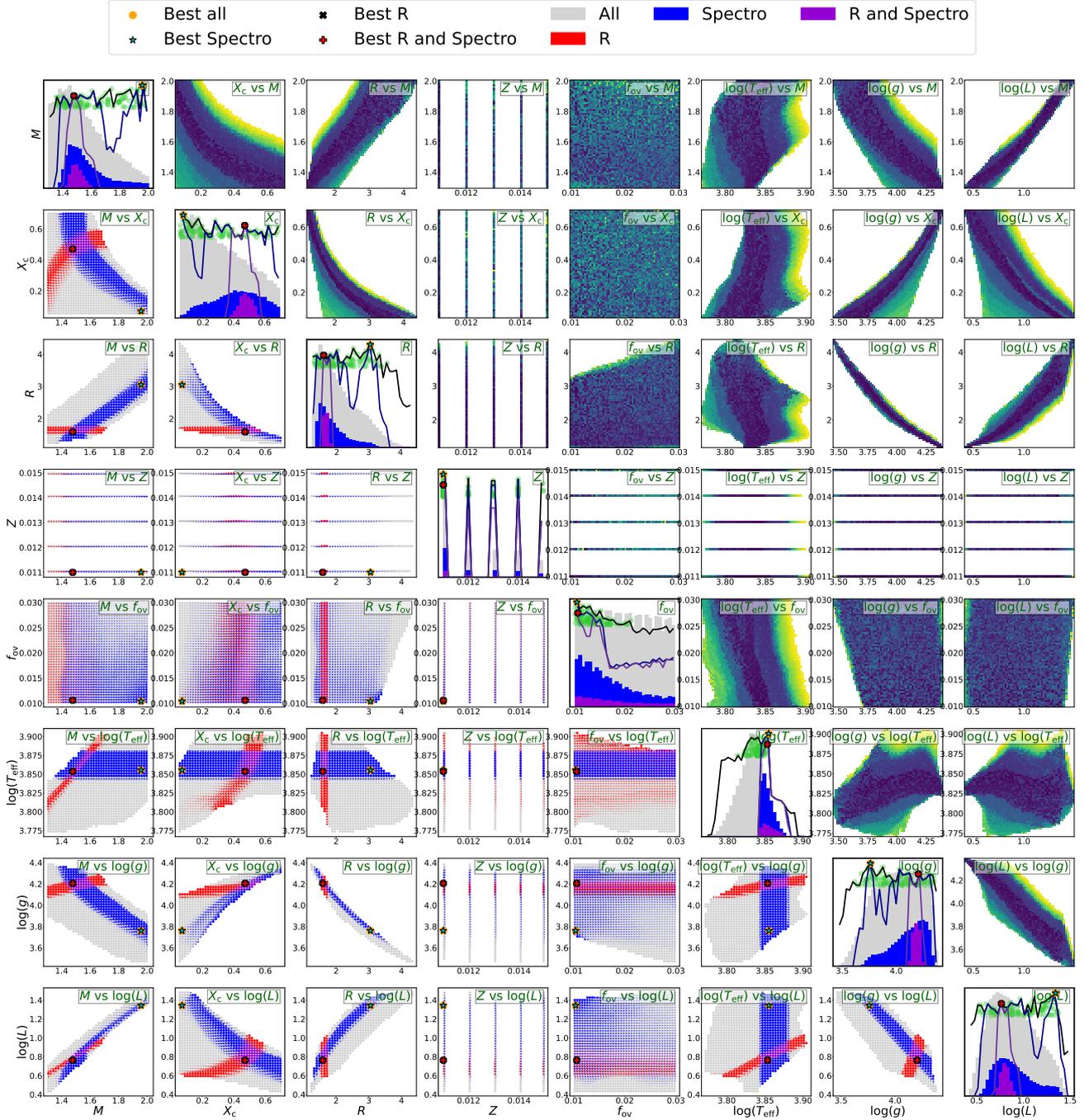}
       \caption{\patli\ modelling results. The main diagonal (emphasised with thicker panel outlines) shows the $1/\chi^2$ envelopes (the maximum value of $1/\chi^2$ per bin, solid lines) and histograms of the $1/\chi^2$ values for each variable along the x-axis when considering purely the asteroseismic fit (All, grey), when only considering models that fall within spectroscopic constraints on \logg\ and \teff\ (Spectro, blue), and when considering only models that satisfy both the spectroscopic constraints and constraints on the stellar radius from eclipse modelling (R and Spectro, purple). Best fitting models are shown for each case. The green markers indicate the 100 best-fitting models. The lower triangular panels are `scatter-pies' (see main text) showing how the different external constraints intersect along each 2D projection of the parameter space, with the best-fitting models also shown. The upper triangular panels show the maximum value of $1/\chi^2$ for each 2D bin (logarithmic scale, with dark colouration indicating a better fit).}
    \label{fig:monsterli}
\end{figure*}

\begin{figure*}
\centering
   \includegraphics[width=1.9\columnwidth]{pics/highres/conservative/conservative_monstrosity_chi2_hist_pie.pdf}
       \caption{\patpes\ modelling results.}
    \label{fig:monsterpes}
\end{figure*}

\begin{figure*}
\centering
   \includegraphics[width=1.9\columnwidth]{pics/highres/hopeful/hopeful_monstrosity_chi2_hist_pie.pdf}
       \caption{\patopt\ modelling results}
    \label{fig:monsteropt}
\end{figure*}

\begin{figure*}
\centering
\includegraphics[width=1.9\columnwidth]{pics/highres/sts/sts_monstrosity_chi2_hist_pie.pdf}
       \caption{\patsts\ modelling results}
    \label{fig:monstersts}
\end{figure*}

\subsection{Pattern fits}

The best-fitting models to the relevant period-spacing patterns that satisfy the radius and spectroscopic constraints are included in Fig. \ref{fig:bfcthree}. These best-fitting models fit the overall shape and location of the periods and period-spacing patterns satisfactorily, and subtle structural features in the observed patterns are able to be modelled in many cases. The fact that the fits make no effort to model the stronger dips and features in the observed patterns is reassuring, as it implies a level of resilience of the methodology to un-modelled physics or badly extracted frequencies (the two are functionally identical and often indistinguishable from a modelling perspective). The quality of the fits, including the apathy towards the larger structural features, is very similar when considering the best overall asteroseismic fit without considering external constraints, as opposed to the best constrained fit shown in Fig. \ref{fig:bfcthree}.

An interesting example of the models' ability to reproduce the small-scale structure is the fit to the \patli\ pattern from radial orders 25-29 in Fig. \ref{fig:bfcthreepatli}. The structure here is only slight, but is able to be reproduced very well by the model. What makes this segment particularly interesting is that the same structure reappears in the fits to the other period-spacing patterns, which include a glitch around the spacings of the 25th and 26th radial order modes. This implies that this structure is an inevitable consequence of matching other parts of the pattern well, most likely the robust first segment (radial orders 16-22). This in turn leads to the conclusion that, at least for this segment, the \patli\ pattern is most likely the correct accurate extraction despite our inability to reproduce its structure in any of the other three patterns.

In contrast, the structure of the third segment (the spacings of radial orders 32-36) is consistent between all three patterns where it was able to be extracted (\patli, \patopt, and  \patsts). In all three cases, it is dominated by a large glitch that is not reproduced by our current models. 

Beyond this sequence, only the \patopt\ and \patsts\ patterns are worth discussing, as \patli\ does not extract beyond the third sequence and \patpes\ extracts only the spacing for the 39th radial order mode. Both \patopt\ and \patsts\ include four consecutive modes at high radial order, although the modes extracted differ between the two patterns: \patopt\ includes spacings for the radial orders 43-45, while \patsts\ includes the spacings of radial orders 44-46. The common radial orders between both patterns, $n=44$ and $n=45$, are in agreement and reproduced by their models. However, while the $n=46$ mode is reproduced by the model fitting the \patsts\ pattern, the $n=43$ mode in the \patopt\ pattern is not. Considering its low amplitude, we consider its detection spurious. 


\subsection{Parameter estimates}
\label{sec:results_paraest}
Figs. \ref{fig:monsterli}-\ref{fig:monstersts} show the results of the parameter estimation component of our modelling work when considering our default case of conservative external constraints. The upper triangular panels present heat maps (dark being better fitting) tracking the maximum of the 1/$\chi^2$ merit function considering all samples, irrespective of external constraints. The panels on the main diagonal present distributions and envelopes of the 1/$\chi^2$ merit function for each parameter, as well as the best-fitting models when considering different external constraints (see Table \ref{tab:constraints}). Finally, the lower triangular panels are `scatter-pies'\footnote{Equivalent to 2-D histogram, but rather than colouring by the count instead a pie chart is drawn in each bin accounting for the relative number of qualitatively different data (here the different combinations of external constraints). Code for producing scatter-pies can be found at: \url{https://github.com/Alex162/scatterpie}} showing how the different external constraints (see Table \ref{tab:constraints}) manifest in each 2D projection of the landscape, with the best-fitting models for each external constraint also shown. Further guidance on reading these figures can be found in the figure caption of Fig. \ref{fig:monsterli}, and the parameters of the best fitting models for each pattern and each external constraint are tabulated in Table \ref{tab:bestcthree_highres}.

Before we delve deeper into each parameter, it is informative to make a few general comments about the results obtained for each pattern. 
The first thing to note is that the best-fitting model for the \patopt\ and \patsts\ patterns, while different between each pattern, is the same when considering the asteroseismology in an unconstrained manner and when considering both the radius and spectroscopic constraints. This is not the case for the \patli\ and \patpes\ patterns; in the \patli\ pattern, the best fitting asteroseismic solution satisfies the spectroscopic constraint but not the radius constraint, while for the \patpes\ pattern, the best fitting asteroseismic solution does not satisfy any of the external constraints. Although this could be coincidence, this can be easily understood by considering that the \patopt\ and \patsts\ patterns contain the most identified spacings, and are therefore more constraining than the shorter \patli\ and \patpes\ patterns. 


The other general comment to be made is that in every case, the best-fitting model from the subset that satisfies the relevant radius constraint (either the tight constraint or the more conservative default constraint) also satisfies the spectroscopic constraints on \teff\ and \logg. This is in spite of the fact that, as can be seen from the lower triangular panels of Figs. \ref{fig:monsterli}-\ref{fig:monstersts}, for many parameters the radius and spectroscopic external constraints have near-orthogonal intersections.

We will now discuss each parameter, beginning with those for which calculations can be made can be made without relying on grids of stellar models (\rotfreqomegacore\ and $\Pi_0$), followed by the model-dependent parameters that are directly obtained from the modelling framework ($M$, \xc, \Z, \fov, \teff, \logg, and \logl) and finally the stellar age, which we estimate from the underlying training sets, and the stellar radius, which we calculate for each model using the stellar mass and its surface gravity.

\subsubsection{Near-core rotation frequency}
\label{sec:results_rotfreq}

As previously explained, the near-core rotation frequency and buoyancy travel time can be directly obtained from the period-spacing patterns without comparison with detailed stellar pulsation models by using a purely theoretical framework. Under the assumptions described in Section \ref{sec:meth:ps_id}, theoretical patterns depending on \rotfreqomegacore\ and $\Pi_0$ were computed using \amigo\ and fit to the observed period-spacing pattern, providing both best-fitting values and an uncertainty estimate for these two parameters. The best-fitting asymptotic patterns to each observed period-spacing pattern can be found in Fig. \ref{fig:patshared} and the lower panels of Figs. \ref{fig:patli2020}-\ref{fig:patsts}.
For the \patli\ pattern, we obtain an estimate for the near-core rotation rate of \rotfreqomegacore\ = $1.562 \pm 0.016 $~\perd, while the other three patterns give estimates of approximately $1.577 \pm 0.010$~\perd. These results are consistent with the $1.58\pm0.01$ \rotfreqomegacore\ estimate from \cite{li2020binary}.

Previously, in Section \ref{sec:lit_constraints}, we estimated the surface rotation frequency for Aa using the radius and radial velocity estimates from \cite{kemp2024eclipse} to be \rotfreqomegasurf\ = $1.54\pm 0.10$ \perd.
Using our \rotfreqomegacore\ estimate of $1.58\pm0.01$, we can conservatively estimate the surface-to-core rotation fraction as: \rotfreqomegasurf / \rotfreqomegacore\ = $0.975\pm 0.064$.
This is consistent with the picture of quasi-rigid core-to-surface rotation presented in Fig. 6 of \cite{aerts2021} for main sequence F-type stars. Aa's \rotfreqomegacore\ of approximately 1.58~\perd\ (18 $\mu \rm Hz$) and \logg\ of $4.14\pm 0.18$ \citep{kemp2024eclipse} place it in the middle of the well-populated clump of other rapidly rotating F-type stars.

As a final comment, we note that it is possible to relax the assumption of rigid rotation, and instead consider a theoretical asymptotic pattern accounting for slightly radially-differential rotation within the star \citep{ogilvie2004,mathis2008,mathis2009,vanreeth2018}.
However, inferring  differential rotation in stellar interiors in this way requires either rotational mode-splittings on top of a period-spacing pattern
(e.g., as in \citealt{triana2015,schmid2016}) or the characterisation of multiple period-spacing patterns within the star to break the degeneracy with rigid rotation \citep{vanreeth2018}. In \cite{vanreeth2018}, it was concluded that for all but the most extreme cases of differential rotation, only stars exhibiting period-spacing patterns in both prograde-dipole and Rossby modes have sufficient distinguishing power to unravel differential from rigid rotation. In KIC~4150611, we identified neither rotational mode splitting -- even in the high-amplitude p-modes -- nor a reliable\footnote{In Appendix \ref{sec:other_freq_groups} we discuss the possibility of a very short $l=2$ series of g-modes.} additional period-spacing pattern, precluding further conclusions of the level of differential core-to-envelope rotation other than to say that the rotation is rigid to within the measurement errors when comparing the asteroseismic near-core rotation and the surface rotation derived from spectroscopy.


\subsubsection{Buoyancy travel time}

Similarly to \rotfreqomegacore, the buoyancy travel time $\Pi_0$ can be estimated directly from \amigo's fit to the pattern using the TAR (see Section \ref{sec:meth:ps_id}). For all patterns except the \patli\ pattern, \amigo\ estimates $\Pi_0$ to be approximately $4024\pm74$~s; for the \patli\ pattern, similarly to \rotfreqomegacore, a slightly lower $\Pi_0$ with a higher uncertainty is obtained ($\Pi_0=3941\pm112$~s). These estimates are consistent with the $\Pi_0$ estimate of $4050\pm80$~s from \cite{li2020all}. We return to the question of the stellar age later, but if Aa was very young (as suggested by \citealt{heliminiak2017}'s isochrone fitting) we would expect a significantly higher $\Pi_0$ ($\Pi_0>4400~s$).


\subsubsection{Stellar mass}

Due to the importance of stellar mass in determining so much of stellar evolution, it is of particular interest to obtain an estimate. Obtaining a precise constraint is difficult in \gdor\ stars due to degeneracies with \xc\ and \Z\ \citep{mombarg2019}. There is indeed significant variation in this parameter when considering the best-fitting models for each pattern and the different external constraints (see Table \ref{tab:bestcthree_highres}).

Considering first the pure asteroseismology, (the `All' case, where all models are considered), we can see that the envelope for the stellar mass is quite flat for all patterns, implying poor constraining power. This is reflected in the uncertainty estimates, which span the entire \gdor\ region. Considering only the best-fitting values, the less constraining \patli\ and \patpes\ patterns arrive at best-fitting models with relatively high mass, 1.96 M\solar\ and 1.69 M\solar\ respectively. The more constraining \patopt\ and \patsts\ models arrive at lower mass best-fitting models, at 1.506 M\solar\ and 1.512 M\solar\ respectively. It is clear from the envelopes that obtaining a confident mass estimate from asteroseismic observables alone for KIC~4150611 is impossible even for the longer period-spacing patterns.

Examining the lower triangular panels of Figs. \ref{fig:monsterli}-\ref{fig:monstersts}, it is clear that in isolation the radius and spectroscopic constraints each provide little information on the mass, with each spanning most of the considered parameter-space. When considered together, however, a near-orthogonal intersecting region is produced that, even with our conservative (2-$\sigma$) constraints  on the stellar radius, \teff, and \logg, constrains the stellar mass to between 1.4 and 1.6 M\solar. Enforcing the tight radius constraint (1.64$\pm 0.01$ R\solar) offers only a slight improvement.

When considering all observables together, the best-fitting models for all period-spacing patterns have masses between 1.47 and 1.51 M\solar, with \patli\ and \patpes\ favouring a lower mass estimate around 1.47-1.48~M\solar\ and \patopt\ and \patsts\ favouring a slightly higher mass around 1.50-1.51~M\solar. The medium resolution sampling arrives at very similar results: 1.46-1.51 M\solar, and the same bifurcation between the \patli/\patpes\ and \patopt/\patsts\ patterns. Typical $1-\sigma$ error estimates are approximately $\pm0.05$ M\solar\ when considering only the sub-sample of models satisfying the external constraints. Enforcing a tight radius constraint changes little, with the various patterns still arriving at best-fitting stellar models with stellar masses between 1.47 and 1.50 M\solar, while the bifurcation between the \patli/\patpes\ and \patopt/\patsts\ solutions disappears. Considering the variation in the best-fitting models, we arrive at a precision of better than 2\% in mass.


\subsubsection{Core hydrogen fraction}

The core hydrogen fraction is a proxy for the stellar age, a property of particular interest for KIC~4150611. \cite{heliminiak2017} estimate the age of the B binary (two G stars) to be approximately 35 Myr from isochrone fitting. Considering the large number of eclipses in an otherwise uncrowded field and the tentative dynamical association between the A triple and the B binary \citep{heliminiak2017}, the co-evolution assumption is quite well motivated for KIC~4150611. Under this assumption, an asteroseismic estimate of \xc\ parameter can be used to put this previous age to the test. A 35 Myr age for KIC~4150611 would imply that Aa is the youngest \gdor\ star observed to date. We will return to the stellar age after concluding our discussion on \cthree's directly modelled parameters.

Firstly, it is important to note that the stellar mass and \xc\ are highly correlated in the asteroseismic fits, reflected in the strong banding present in the 2D envelope seen in the upper triangular panels showing \xc\ vs \M\ in Figs. \ref{fig:monsterli}-\ref{fig:monstersts}. For this reason, much of the previous discussion surrounding \M\ is relevant to \xc. 

Once again first considering the pure asteroseismic estimations, due to the correlated nature of \xc\ and \M\ the high mass estimates of the \patli\ and \patpes\ patterns translate to low estimates of \xc, with the converse true for the \patopt\ and \patsts\ patterns. Further, we note that once again the envelope is mostly flat, and therefore poorly constraining, with the $1-\sigma$ error estimate once again spanning almost the entire range of possible values of \xc\ (0-0.7).

Considering the external spectroscopic and eclipse modelling constraints without the benefit of asteroseismology, a fairly broad bounding constraint between 0.38 and 0.58 in \xc\ can be placed\footnote{The initial core hydrogen fraction is approximately 0.7, with the precise value depending on the metallicity}. 

Considering the external constraints and the asteroseismology in conjunction, we see that the best-fitting models have \xc\ varying between 0.42 and 0.49 from the high resolution sampling, with \patli\ and \patpes\ favouring higher \xc\ and \patopt\ and \patsts\ favouring lower \xc, as expected given their preferences towards a higher stellar mass estimate. Estimated $1-\sigma$ uncertainties for these constrained values are at most $\pm0.1$, and appear to be significantly lower for some patterns ($\pm0.04$ in the case of \patsts, for example). Enforcing the tight radius constraint results in best values for \xc\ between 0.42 and 0.44.

\subsubsection{Metallicity}

The sampling in \Z\ is low resolution, with only 5 different cases able to be considered over a narrow range (\Z=0.011-0.015). There does, however, seem to be a preference towards lower metallicity solutions, especially when the radius and spectroscopic constraints are taken into consideration, although the envelope is very flat.

From atmospheric analysis of the disentangled spectrum of Aa in \cite{kemp2024eclipse}, we have a metallicity estimate of [M/H]=$-0.23\pm0.06$, corresponding to \Z=$0.0084\pm 0.0011$ (using the solar metallicity of \Z\solar=0.0142857 from \citealt{asplund2009}, following \citealt{mombarg2021}). This places the star in the middle of the metallicity range of \gdor\ stars with metallicity measurements from high-resolution spectroscopy \citep{gebruers2021}. This metallicity is slightly outside the bounds of the training set for \cthree, which has a lower limit of 0.011. Armed with the knowledge of the spectroscopic metallicity measurement, we note that the preference towards lower metallicity could imply a small degree of sensitivity to \Z\ in the asteroseismic fits. However, precise estimation of metallicity from g-modes in isolation is unlikely to be possible in practise due to degeneracy between the stellar mass and metallicity.

\subsubsection{Exponential overshooting factor}

The exponential overshooting factor determines the degree of exponential core overshooting in the core boundary region. \cite{mombarg2021} vary this from between 0.01 and 0.03 to form their training set.

For all patterns except \patopt, which has a (very flat) `U' distribution spanning the full range of \fov, there is a clear preference towards low exponential overshooting. This is consistent with the empirical \M-\fov\ relation of \cite{claret2017}, who found values of \fov\ between 0.005 and 0.02 across the \gdor\ range, with lower values associated with the low-mass end regime 1.4-1.5~M\solar. We note that \cite{mombarg2024} find no such correlation between the stellar mass and \fov, but do find that low values of \fov\ are most probable in \gdor\ stars.

Despite the fact that \fov\ can affect the stellar radius and other external properties such as the luminosity, we find that the radius and spectroscopic constraints provide no useful constraint on this parameter either when considered in isolation or their intersection.

\subsubsection{Effective temperature and surface gravity}

From atmospheric analysis on the disentangled spectra using \gssp\ \citep{tkachenko2015}, we have an estimate for the effective temperature and surface gravity of \teff=$7280\pm70$~K (log(\teff)=$3.8621\pm0.0042$~dex) and \logg=$4.14\pm0.18$~dex. 

A centrally peaked envelope around (approximately) the spectroscopic solution is present with or without external considerations for all patterns, although the \patopt\ pattern's envelope is significantly flatter and broader. The best-fitting log(\teff) solutions -- excluding the extreme case of log(\teff)=3.843~dex for the \patpes\ pattern considered without external constraints -- vary from between 3.855-3.856~dex, placing them slightly below the lower bound of the 1-$\sigma$ uncertainty of the \gssp\ estimate. Considering that the effective temperature was not part of the merit function, this level of agreement is quite good, although we note that the best-fitting models for all patterns slightly under-predict the spectroscopic solution. Uncertainty estimates for the upper limit of log(\teff) can be as high as $+0.008$~dex, and are even larger when considering lower temperatures (approximately $-0.012$~dex). These uncertainties are sufficiently large that they overlap with the spectroscopic estimate for \teff. 

Spectroscopy only places loose constraints on \logg\ (\logg=$4.14 \pm 0.18$~dex, \citealt{kemp2024eclipse}), with the 2-$\sigma$ uncertainty region essentially spanning the entire \gdor\ range. However, the radius constraint from the eclipse modelling narrows the viable region considerably. Combining the spectroscopic and radius constraints allows bounding constraints to be placed on \logg\ of 4.15-4.25~dex, consistent with the spectroscopic solution. Within this far smaller region, the envelope of the asteroseismic merit function is peaked, although the best-fitting solution still varies significantly from pattern to pattern. \patli\ and \patpes\ arrive at higher estimates for \logg, 4.206~dex and 4.223~dex, respectively, while the best fitting solutions for \patopt\ and \patsts\ arrive at \logg\ of 4.168~dex and 4.176~dex, respectively. For all patterns, a slightly higher \logg\ than the face value of the spectroscopic solution is preferred, although all three cases are, as a result of the already constraining intersection between the radius and spectroscopic solutions, well within the spectroscopic 1-$\sigma$\ uncertainty region. Typical 1-$\sigma$ uncertainties of the constrained asteroseismic modeling are estimated to be approximately $\pm 0.04$~dex.

\subsubsection{Luminosity}

The luminosity $\chi^2$ envelope is, similarly to \logg, characterised by a broad and relatively flat envelope when considering the asteroseismology in isolation. The spectroscopic constraints affect this little, although it is interesting to note that, when looking at the histogram of the asteroseismic merit function, the peak roughly coincides with the region consistent with the intersection between the spectroscopic and radius constraints.

Also similarly to \logg, imposing both the spectroscopic and radius constraints allows bounding constraints to be placed on \logl\ (0.7-0.9 dex). \patli\ and \patpes\ arrive at lower estimates for \logl\ of 0.763~dex and 0.745~dex respectively, while the best fitting solutions for \patopt\ and \patsts\ arrive at estimates for \logl\ of 0.819~dex and 0.809~dex respectively. Estimated $1-\sigma$ uncertainties are approximately $+0.006$ and $-0.003$ for the externally constrained asteroseismic solutions.

From \emph{Gaia} DR3, we have log($L$) estimates for Aa of log($L$) = $0.804\pm0.017$~dex assuming a light fraction of 0.82, and log($L$) = $0.838\pm0.017$~dex assuming a light fraction of 0.92 (see Section \ref{sec:lit_constraints}).
Considering our best-fitting solutions, the lower values of the \patli\ and \patpes\ solutions for \logl\ (0.763~dex and 0.745~dex respectively) are closer to, but still underestimate, even the lower light fraction (0.85) luminosity estimate preferred by the eclipse modelling and \cite{heliminiak2017}. The higher values of \logl\ implied by the best-fitting solutions of the \patopt\ and \patsts\ patterns (0.819~dex and 0.809~dex), however, are consistent with the \emph{Gaia} luminosity regardless of which light fraction is assumed.

\subsubsection{Stellar age}

The stellar age can be estimated from the mass and core hydrogen fraction, although there are secondary physical influences such as birth composition, rotation, and degree of core-envelope mixing. The stellar age is not directly predicted by \cthree, but an estimate can be obtained by searching within the \cthree\ training set for the nearest model in terms of \M, \Z, and \fov, where mass is prioritised, and then interpolating the age from the evolution history using \xc. The resolution within the training set is too poor for any useful inference of most other stellar properties, but the dominant dependence on \M\ and \xc\ makes this technique viable for the stellar age. We provide tabulated information on the nearest training model for each solution in the online supplementary material, but caution against using it to infer other stellar properties.

Doing so results in stellar age estimates of 830 Myr, 1230 Myr, 1100 Myr, and 1070 Myr for the \patli, \patpes, \patopt, and \patsts\ patterns respectively for the pure asteroseismic best-fitting solutions. The externally constrained solutions have age estimates of 1280, 1200, 1100, and 1070 Myr for the \patli, \patpes, \patopt, and \patsts\ patterns, respectively.

This is clearly inconsistent with the very young 35 Myr age estimate for the B binary from \cite{heliminiak2017}. Such a young age would require a very high \xc\ estimate for any star within the \gdor\ range; even a 2 M\solar\ star has a main sequence lifetime of almost a 1000 Myr.

As previously discussed, the external constraints -- which are quite conservative -- impose a maximum value for \xc\ of 0.58 when considering their intersection, already resulting in stellar ages significantly older than 35 Myr. Considering the asteroseismology in isolation, there is little evidence of a peak at very high \xc\ when considering the envelope, and none when external constraints are considered. Further, the buoyancy travel time is too low for the star to be so young. The approximately 100 Myr-old \gdor\ stars in NGC 2516 have buoyancy travel times of around 4800~s \citep{li2024cluster}, for example, and these may be the youngest \gdor\ stars aged to date.

In \cite{kemp2024eclipse}, radius and mass ratio estimates were obtained not only for the primary pulsator Aa, but also the members of the tight Ab eclipsing binary, composed of two K/M dwarfs. The radius estimates for these two dwarf stars were significantly smaller than the radii implied by \cite{heliminiak2017}'s isochrone fits to the B binary, which placed these stars on the pre-main sequence. Their smaller size, therefore, implies older stars have already contracted to the zero-age main sequence. \cite{kemp2024eclipse} noted that this could imply that the co-evolution assumption between the A triple and the B binary could be invalid, but could also simply be due to the large inherent uncertainty associated with isochrone fits. The \cite{heliminiak2017} isochrone fit for Aa, a 1.64 M\solar\ star with a radius of only 1.376 R\solar, was also significantly different than the approximately 1.65 R\solar\ radius estimate from the eclipse modelling, which could also imply an older star. Considering all of this information, we consider an age estimate of $1100\pm100$ Myr to be a robust update on the previous isochrone-based age estimate.


\subsubsection{Stellar radius}

The stellar radius of Aa and its uncertainty is dealt with in detail in \cite{kemp2024eclipse}, and forms one of the external constraints. Here, we discuss the ability of the asteroseismic and spectroscopic observables to estimate the stellar radius and the level of agreement between the modelling work and the radius estimate from \cite{kemp2024eclipse}.

Considering the asteroseismology in conjunction with the spectroscopic constraints, a bimodal envelope pattern structure appears, with one broad peak at low stellar radius (consistent with the eclipse radius) and a second at high radius (around 3 R\solar.) This feature is present (although varies in prominence) in all four considered patterns.

Considering the level of consistency with the radius estimate from \cite{kemp2024eclipse}, the best-fitting asteroseismic models with radius and spectroscopic external constraints applied have radii of 1.59, 1.55, 1.68, and 1.66 R\solar\ for the \patli, \patpes, \patopt, and \patsts\ patterns respectively in the high-resolution sampling. The $1-\sigma$ uncertainties are estimated to be as high as $\pm0.1$ R\solar, with the lower \patli\ and \patpes\ estimates being skewed towards higher radius estimates.\cite{kemp2024eclipse} considered the effect of variation in the light-fraction of Aa, finding \RAa$=1.65\pm0.01$~R\solar, \RAa$=1.62\pm0.01$~R\solar, and \RAa$=1.61\pm0.02$~R\solar, for the 0.92-0.96, 0.83-0.87, and `free' light fraction cases respectively. On the basis of their spectroscopic analysis, \cite{kemp2024eclipse} preferred the 0.92-0.96 light fraction solution. In the context of our asteroseismic modelling, all three of \cite{kemp2024eclipse}'s radius estimates are so similar that it would be a stretch to say one is preferred. We do note, however, that the more constraining asteroseismic patterns, \patopt\ and \patsts, both arrive at best-fitting values of the stellar radius (1.68 R\solar\ and 1.66 R\solar) that are consistent with the radius estimates from the eclipse modelling.

\subsection{MCMC grid search}

Five different grids of stellar models (grids A-E, see section \ref{sec:meth:mcmc}) were considered for as part of the MCMC parameter-based grid-search. All MCMC simulations converged to a solution, but not all solutions were able to recover all input parameters satisfactorily, implying that a consistent model did not exist within the grid. Grids A (a \Z=0.011, non-rotating grid from \cite{mombarg2021}), C ( a solar metallicity, rotating grid from \cite{mombarg2024gaia}), and E (a \Z=0.008, rotating grid with equivalent physics to \cite{mombarg2024gaia}) were able to recover all input parameters, and we will discuss their solutions later in this section. Grids B (a solar metallicity, rotating grid from \cite{mombarg2024} including hydrodynamic rotational mixing) and D (a \Z=0.0045, rotating grid from \cite{mombarg2024gaia}) were not able to recover the input parameters.

\begin{figure}
\centering
   \includegraphics[width=1\columnwidth]{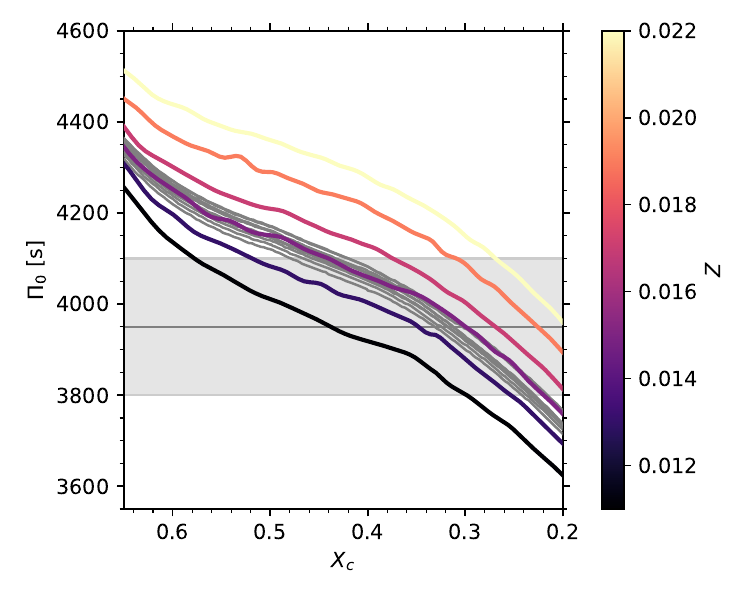}
       \caption{Buoyancy travel time vs central H fraction for a 1.5 M\solar\ star, for various metallicities, using data from \cite{mombarg2021}. The buoyancy travel time for KIC~4150611 (see Section \ref{sec:results}, which is an input for the MCMC grid-search component of the modelling, is shown as the horizontal grey band. The grey lines are the solar metallicity, rotating stellar models from \cite{mombarg2024} (grid B); the width of the band that they form is indicative of the variation due to stellar rotation.}
    \label{fig:pi0_xc_z}
\end{figure}

Grid B has the most sophisticated stellar physics treatment for stellar mixing; however, its computation at solar metallicity poses an issue. Figure \ref{fig:pi0_xc_z} shows the large effect that metallicity has on the crucial asteroseismic parameter $\Pi_0$, on which the ageing of the star chiefly depends. Comparing grids A and B, we found the effect of metallicity on the buoyancy travel time was at least several hundred seconds, compared to only a few tens of seconds from the rotation. The metallicity of KIC~4150611 is determined spectroscopically to be \Z\ = 0.0084, significantly lower than the $Z \approx 0.014$ solar metallicity adopted in \cite{mombarg2021,mombarg2024}. This miss-match in metallicity may be responsible for the inability of the MCMC simulation using grid B to find a solution consistent with all observables.

However, the treatment of mixing can also affect $\Pi_0$ significantly; the solar metallicity models from \cite{mombarg2024gaia} (grid C) have a similar $\Pi_0$ profile to the $Z=0.011$ profile in Figure \ref{fig:pi0_xc_z} (grid A), differing significantly from the solar metallicity models of \cite{mombarg2024} (grid B). This implies that internal physical choices can be as important as matching the metallicity when using $\Pi_0$. The $Z=0.0045$ (grid D) models from \cite{mombarg2024gaia} have significantly lower values of $\Pi_0$, as anticipated, but also a very flat $\Pi_0$-\xc behaviour between \xc\ = 0.55 and \xc\ = 0.3, implying lower sensitivity to stellar ages in this region. This flattening is particularly pronounced for higher \fov.

Considering now only the MCMC simulations where the input parameters were able to be recovered within uncertainty (grids A, C, and E), we find that the results are broadly consistent with the pattern modelling approach using \cthree\ and consistent with each-other. We also note that, as we might expect, grid E -- which most closely matches the metallicity of KIC~4150611 -- does a slightly better job at recovering the input parameters.

From our MCMC modelling, we estimate KIC~4150611 to be a $1.52\pm0.06\$$M\solar\ star with $X_c=0.48\pm0.08$, values consistent with those found by our \cthree\ modelling, albeit slightly higher in the case of $X_c=0.48\pm0.08$. This methodology was unable to constrain \fov, likely due to this parameter being insufficiently constrained by $\Pi_0$ alone. This was also the case in \cite{fritzewski2024}, and highlights the added value of pattern-modelling relative to parameter-based grid search methods.

However, one advantage of the MCMC method is its ability to sample all evolutionary parameters accessible through the MCMC chain, giving access to stellar properties that cannot be directly accessed via \cthree. We estimate the stellar age to be $1110\pm150$~Myr, in agreement with our previous estimates. Further, we find a convective core mass fraction $M_\mathrm{conv}/M=0.09\pm0.01$, which is consistent with the distribution of other \kepler\ \gdor\ stars \citep{mombarg2019}.

\section{Conclusions}

\label{sec:conc}

In this work, we modelled the gravity-mode period-spacing pattern of the Aa component of KIC~4150611 with the goal of estimating its stellar parameters. This was done with attention to how the different external constraints from spectroscopy and photometric eclipse modelling interplay with the asteroseismic information. We also considered four different period-spacing patterns for the system to account for systematic differences in pattern identification and frequency extraction methods.

We find that pattern-dependent frequency variations are less important than the completeness of the pattern, particularly in terms of how far the pattern extends to high radial orders. In several key parameters, such as the stellar mass and core hydrogen fraction, there is a clear bifurcation between the parameter estimation from the shorter patterns, \patli\ and \patpes, and the longer \patopt\ and \patsts\ patterns. The parameters for the longer \patopt\ and particularly the \patsts\ patterns agree best with external information from the eclipse modelling, spectroscopy, and the system luminosity calculated from \emph{Gaia} DR3 data. 

For patterns of similar length and completeness to KIC~4150611, the pattern may not be  constraining enough for confident conclusions to be drawn for many stellar parameters without the benefit of external constraints. Notable exceptions include near-core properties such as the degree of exponential overshoot and the near-core rotation. The effective temperature also appears to be able to be consistently estimated from the seismology alone, likely due to it being correlated with the buoyancy travel time.

When the external spectroscopic and  constraints are considered in isolation, they are generally uninformative for any given parameter. However, the intersection of their 2-$\sigma$ uncertainty regions does allow for useful -- although imprecise -- bounds to be placed on the mass, central hydrogen fraction, and luminosity. However, it is unhelpful when considering the metallicity, degree of core overshooting, and the near-core rotation.

We find the near-core rotation rate to be $1.58\pm0.01$~\perd, consistent with \cite{li2020binary}. Combined with an estimate for the surface rotation frequency of $1.54\pm0.10$~\perd, this is consistent with near-perfect rigid rotation of the radiative zone. We also estimate the buoyancy travel time to be $4024\pm74$~s, also consistent with \cite{li2020binary}.

Considering the intersection of the 2-$\sigma$ uncertainty regions of the spectroscopic and  constraints in conjunction with the asteroseismology, we arrive at the best motivated constraints of the stellar parameters. The stellar mass we estimate to be between 1.47 M\solar\ and 1.51 M\solar\ ($\pm0.05$ M\solar), preferring the high-mass solutions around 1.51 M\solar\ from the \patopt\ and \patsts\ patterns. The core hydrogen fraction we find to be between 0.42 and 0.49 ($\pm 0.04$). The lower values correspond to the preferred \patopt\ and \patli\ patterns; we also note that core hydrogen fractions no higher than 0.58 are permitted based on the intersecting radius and spectroscopic constraints. Stellar radii we find to be between 1.55~R\solar\ and 1.68~R\solar\ (with uncertainties no higher than $\pm0.1$R\solar), preferring the higher radius solutions from the \patopt\ and \patsts\ patterns, which we note are more consistent with the eclipse modelling from \cite{kemp2024eclipse}. The metallicity shows, for all patterns and constraints, a preference to the lower metallicity bound of \cthree. This is consistent with atmospheric modelling estimates for the metallicity from \cite{kemp2024eclipse}, which place the star slightly below the metallicity range able to be modelled by \cthree. The best-fitting models consistently prefer low values of exponential overshooting regardless of the pattern or constraint considered. This is consistent with the low-mass end of the empirical \M-\fov\ relation for \gdor\ stars put forth by \cite{claret2017}, as well as the probability distributions for \gdor\ stars computed by \cite{mombarg2024}.


The best fitting models have \teff\ between 7160~K and 7180~K, placing them just below the 1-$\sigma$ uncertainty region from the spectroscopic analysis ($7280\pm70$~K). The uncertainty estimates for log(\teff) of approximately $\pm0.008$~dex are sufficient to overlap with the spectroscopic solution. Helped considerably by the intersecting radius and spectroscopic constraints, the best-fitting models have \logg\ varying between 4.168~dex and 4.223~dex ($\pm 0.04$~dex), with the preferred \patopt\ and \patli\ fits arriving at results at the lower end of the range. We find that \logl, also helped considerably by the intersecting spectroscopic and radius constraints, is between 0.745~dex and 0.819~dex ($\pm0.005$~dex), preferring the high luminosity solutions from the \patopt\ and \patsts\ patterns. These higher solutions are consistent with the stellar luminosity calculated from \emph{Gaia} DR3 and light fractions for Aa from \cite{kemp2024eclipse} and \cite{heliminiak2017}.

 We estimate the age of Aa to be approximately $1100\pm100$~Myr, much older than the 35~Myr age implied by \cite{heliminiak2017}'s isochrone fits. This older age from asteroseismology is also supported by the radius measurement from \cite{kemp2024eclipse}, and serves as a reminder of the important role that asteroseismology can play in stellar parameter estimation. 

 From our parameter-based MCMC grid-search, we were able to recover similar solutions for the stellar parameters of KIC~4150611 Aa, including the stellar age. Comparing different model grids, the significant effect of metallicity on the buoyancy travel time was evident. The need for careful treatment of the envelope mixing was also evident, potentially affecting the buoyancy travel by as much as the metallicity.

As future work it would be interesting to revisit the isochrone fitting of the B system in an attempt to confirm its age, ideally in parallel with an independent isochrone aging of the components of the A triple system using their radii and estimated seismic masses. KIC~4150611 is not part of a particularly crowded field, so the fact that there are four sets of eclipses is highly suggestive of a physical association, and likely a co-evolutionary origin, between its A, B, and C components. A revisit of the aging of different system components could provide clarity on this issue, and greatly assist in establishing the evolution history of this fascinating system.

\begin{acknowledgements}
The authors wish to thank Sarah Gebruers, Dominic Bowman, and Annachiara Picco for their useful discussion and input.
The research leading to these results has received funding from the KU\,Leuven Research Council (grant C16/18/005: PARADISE), from the Research Foundation Flanders (FWO) under grant agreement G089422N (AT, CA) and 1124321N (LIJ), as well as from the BELgian federal Science Policy Office (BELSPO) through PRODEX grant PLATO. V.V. gratefully acknowledges support from the Research Foundation Flanders (FWO) under grant agreement N°1156923N (PhD Fellowship). JSGM acknowledges funding the French Agence Nationale de la Recherche (ANR), under grant MASSIF (ANR-21-CE31-0018-02).
CA acknowledges funding by the European Research Council under grant ERC SyG 101071505. Funded by the European Union. Views and opinions expressed are however those of the author(s) only and do not necessarily reflect those of the European Union or the European Research Council. Neither the European Union nor the granting authority can be held responsible for them. This paper includes data collected by the Kepler mission, which are publicly available from the Mikulski Archive for Space Telescopes (MAST) at the Space Telescope Science Institute (STScI). Funding for the Kepler mission is provided by the NASA Science Mission Directorate. STScI is operated by the Association of Universities for Research in Astronomy, Inc., under NASA contract NAS 5–26555.
\end{acknowledgements}

%
%
\bibliographystyle{aa}
\bibliography{bibfile.bib}

\appendix

\section{Detailed period-spacing pattern plots}

\begin{figure*}
\centering
   \includegraphics[width=1.99\columnwidth]{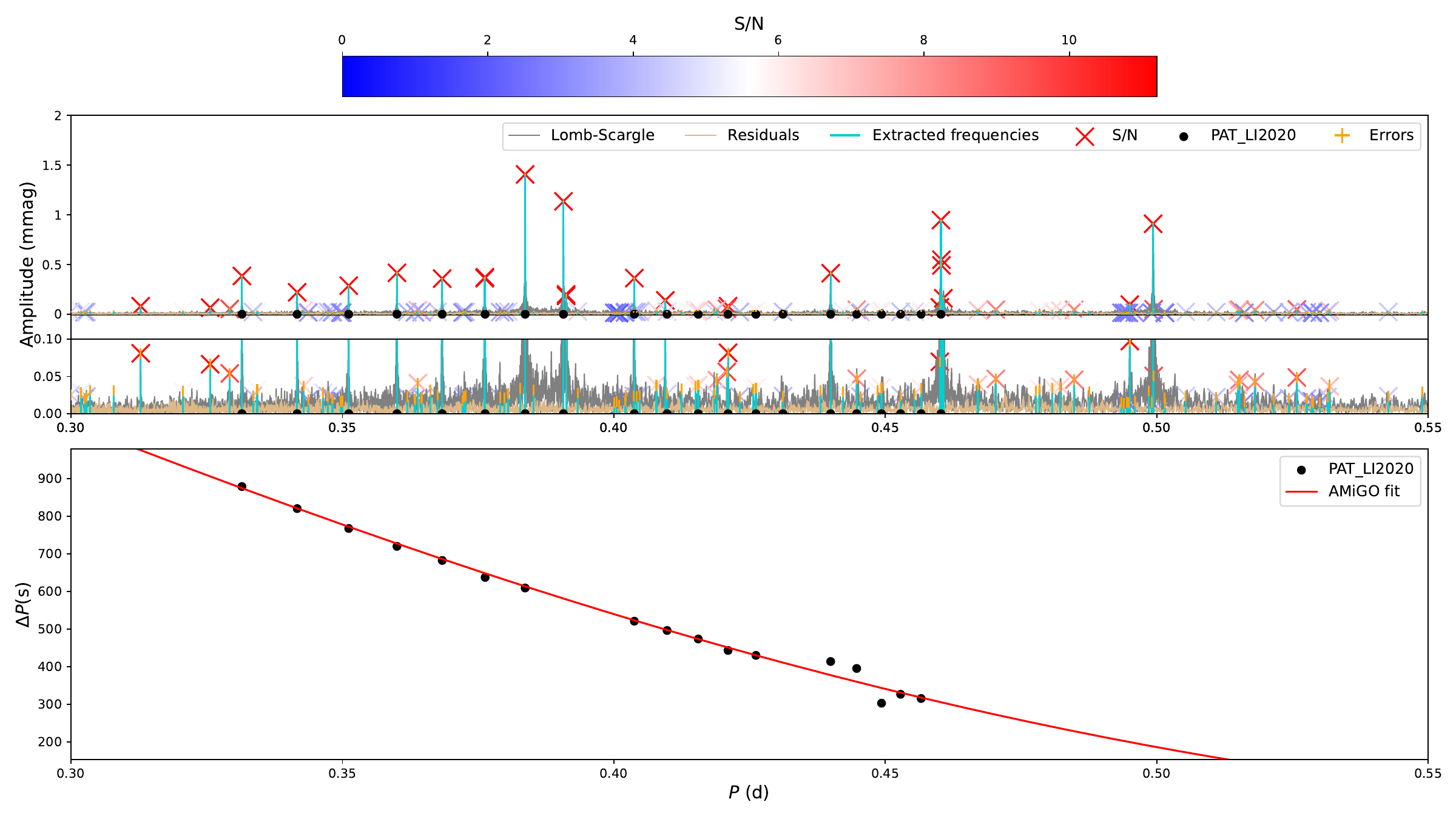}
       \caption{\patli\ period-spacing pattern from \cite{li2020binary}. The Lomb-Scargle periodogram is shown in grey, the extracted frequencies in light blue, and the residual periodogram with the extracted frequencies removed in orange. Orbital harmonics are excluded. The SNR ratio is indicated by the colour of the `X' symbol for each extracted frequency, while the amplitude and period errors (more easily seen in the inset panel, which better shows the low-amplitude behaviour) are shown in orange. The extracted frequencies selected to be part of the pattern are marked with a filled black circle. Note that only the black point and the \amigo\ fit and predictions are specifically related to the \patli\ pattern; the Lomb-Scargle periodogram and its residuals, as well as the extracted frequencies (and their associated SNR and errors), are included from the \pofour\ extraction for comparison purposes.}
    \label{fig:patli2020}
\end{figure*}

\begin{figure*}
\centering
   \includegraphics[width=1.99\columnwidth]{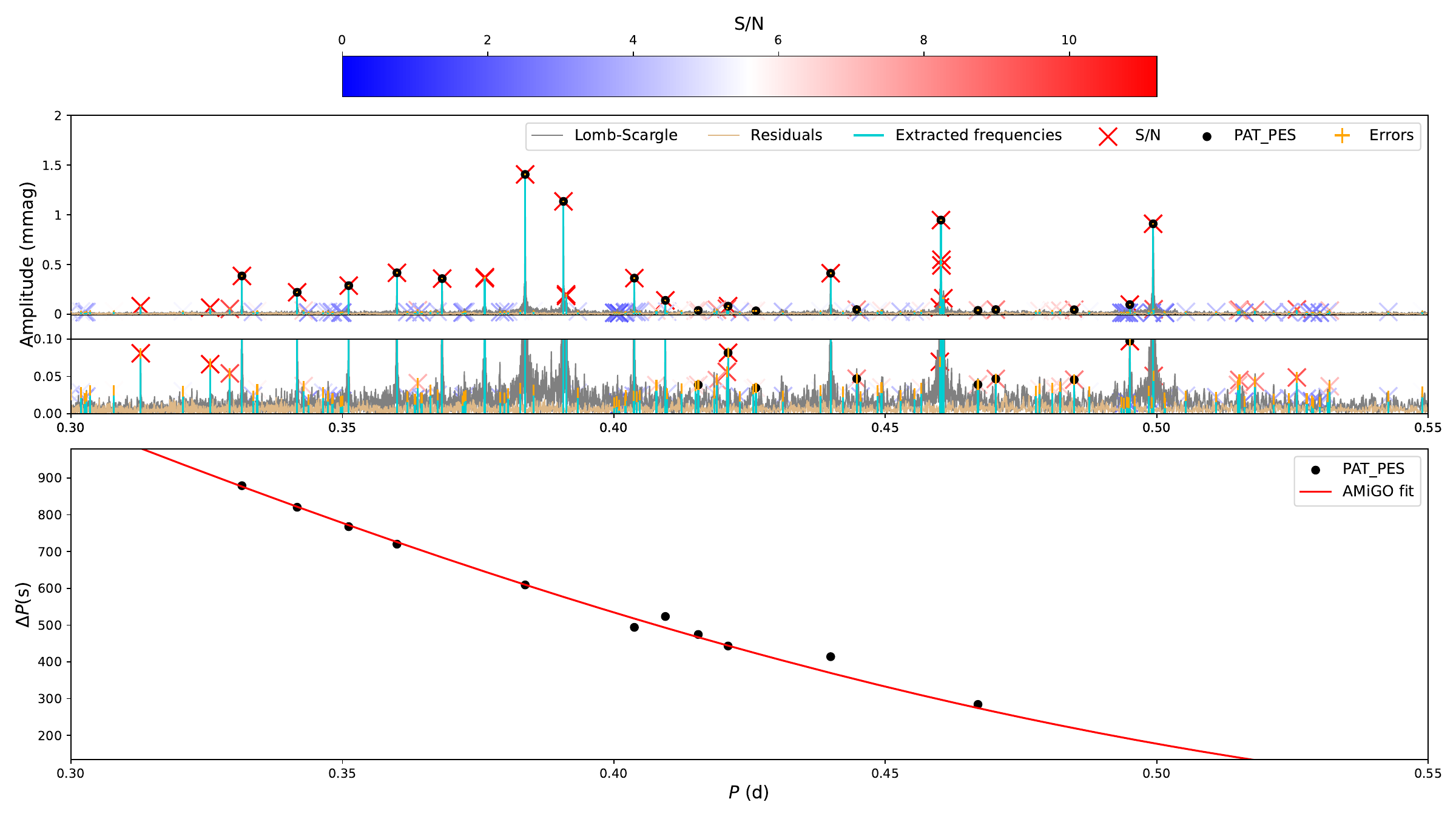}
       \caption{\patpes\ period-spacing pattern. Orbital harmonics are excluded. The high-amplitude period at 0.36d is excluded in this pattern only due to its near-perfect coincidence with an 8.65d orbital harmonic. A detailed description of all symbols and information can be found in Figure \ref{fig:patli2020}.}
    \label{fig:patpes}
\end{figure*}

\begin{figure*}
\centering
   \includegraphics[width=1.99\columnwidth]{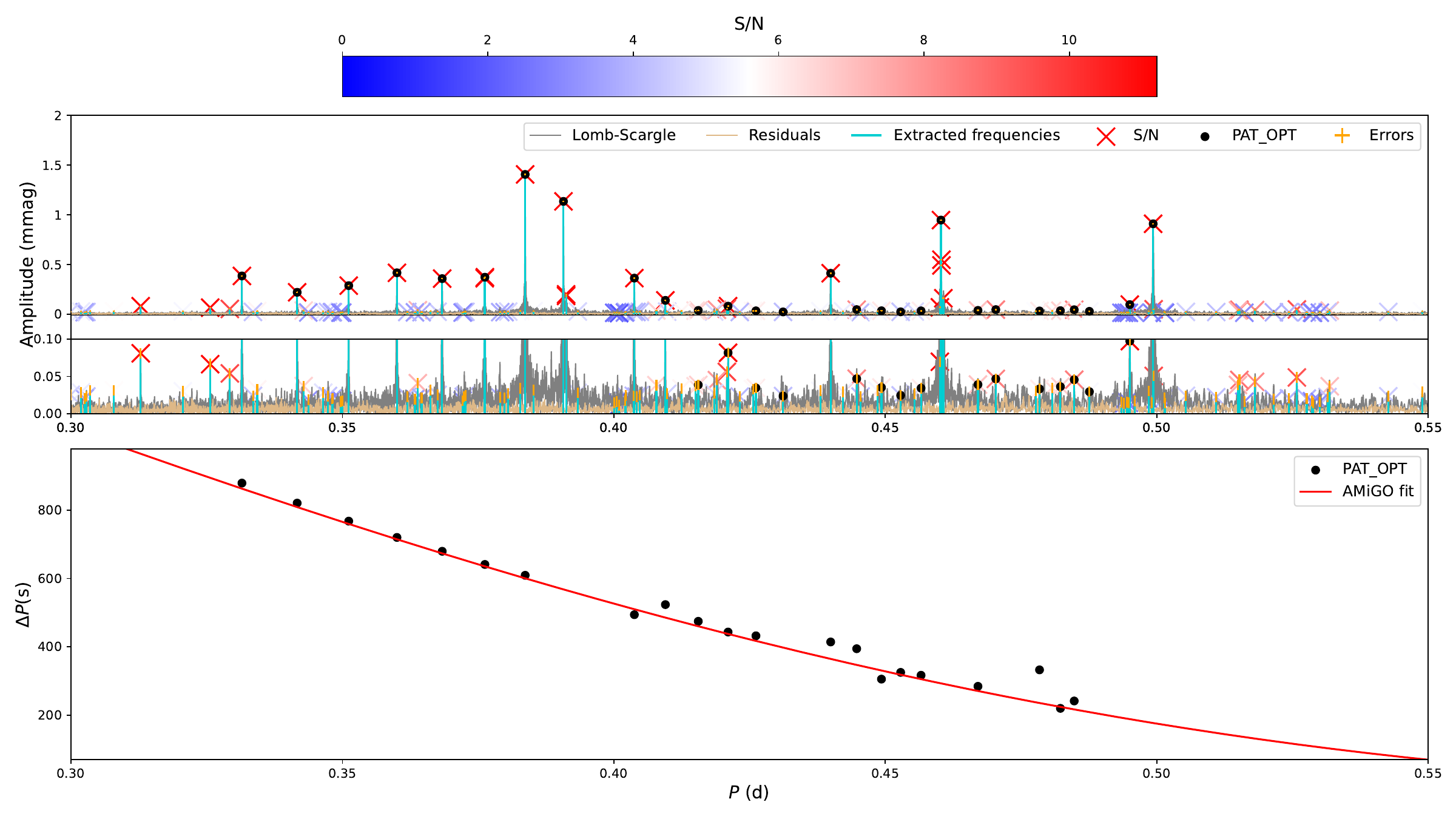}
       \caption{\patopt\ period-spacing pattern. Orbital harmonics are excluded. A detailed description of all symbols and information can be found in Figure \ref{fig:patli2020}.}
    \label{fig:patopt}
\end{figure*}

\begin{figure*}
\centering
   \includegraphics[width=1.99\columnwidth]{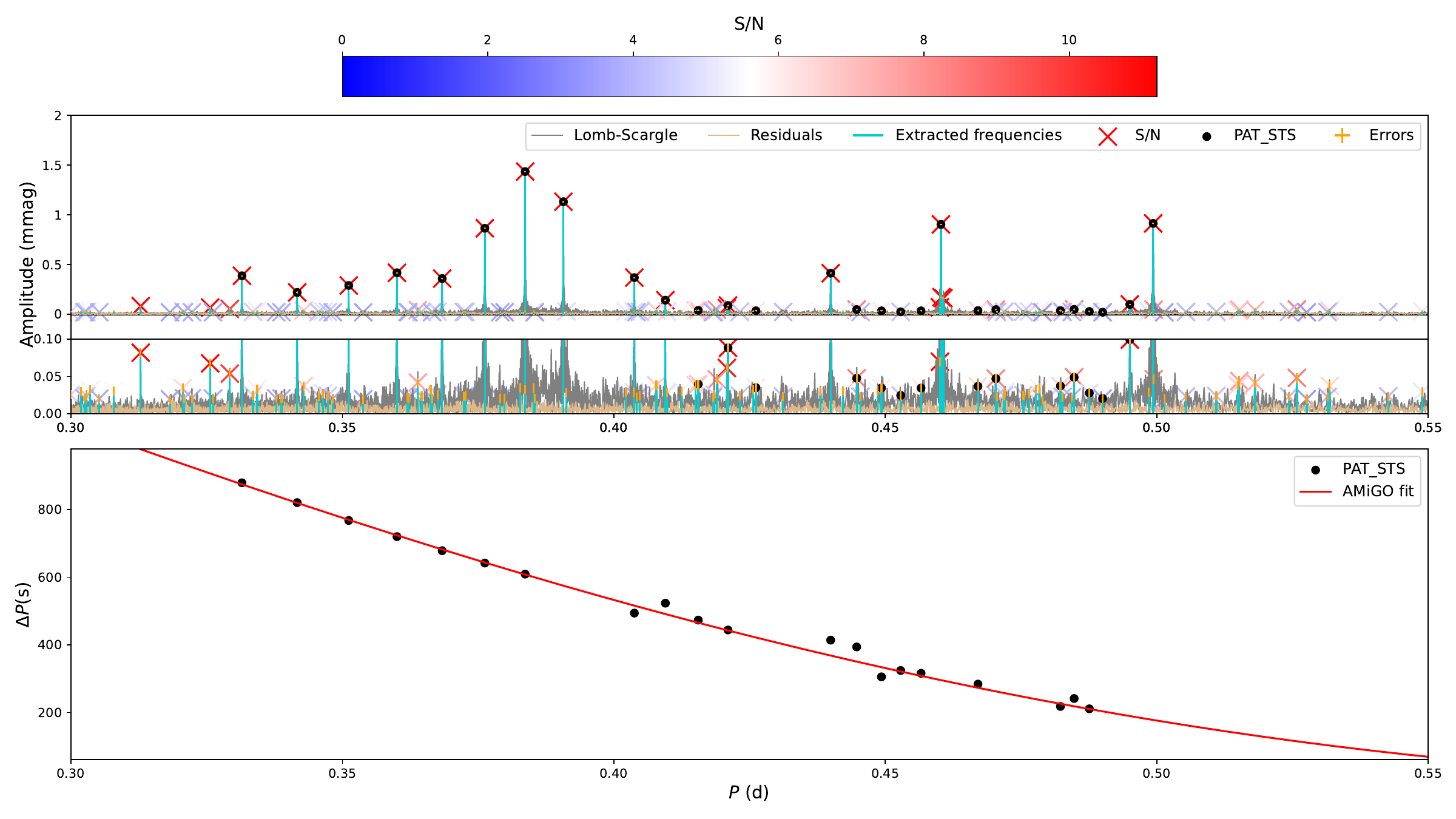}
       \caption{\patsts\ period-spacing pattern. Orbital harmonics are excluded. A detailed description of all symbols and information can be found in Figure \ref{fig:patli2020}.}
    \label{fig:patsts}
\end{figure*}

\section{Tight $R$ constraint}

\begin{figure*}
\centering
\begin{subfigure}{0.5\textwidth}
\centering
\includegraphics[width=1\columnwidth]{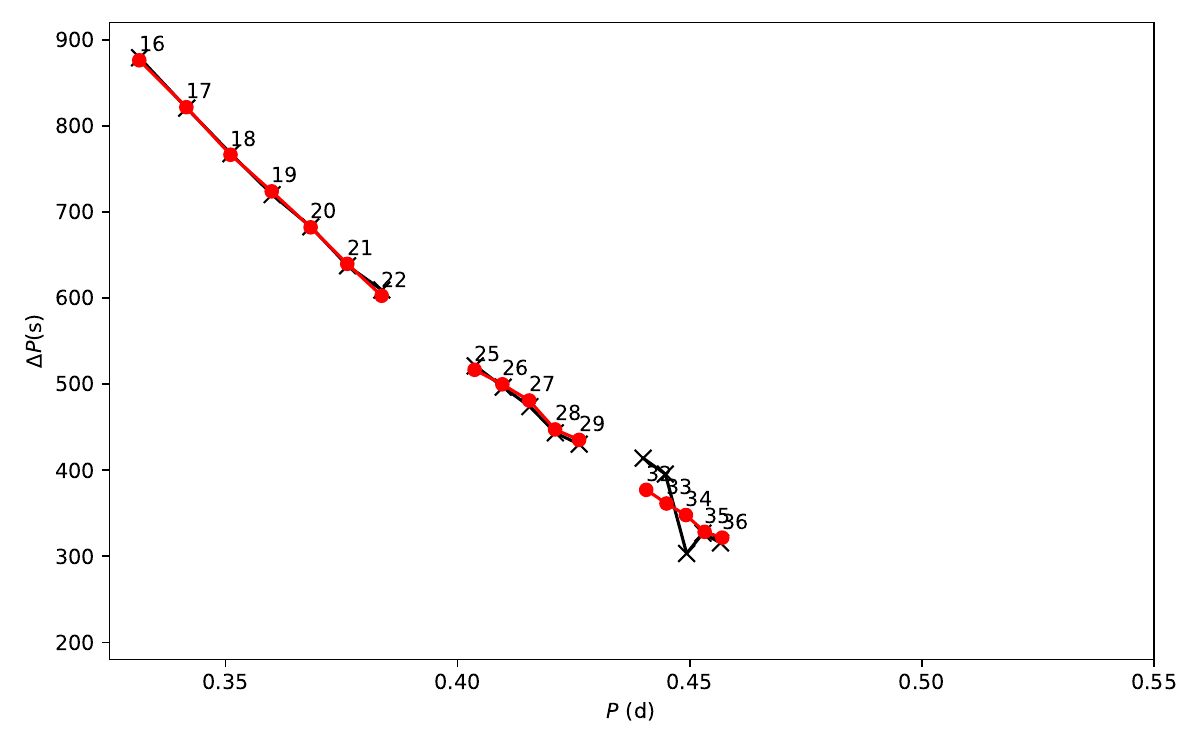}
\caption{\patli}
\label{fig:bfcthreepatlitightr}
\end{subfigure}%
\begin{subfigure}{0.5\textwidth}
\centering
\includegraphics[width=1\columnwidth]{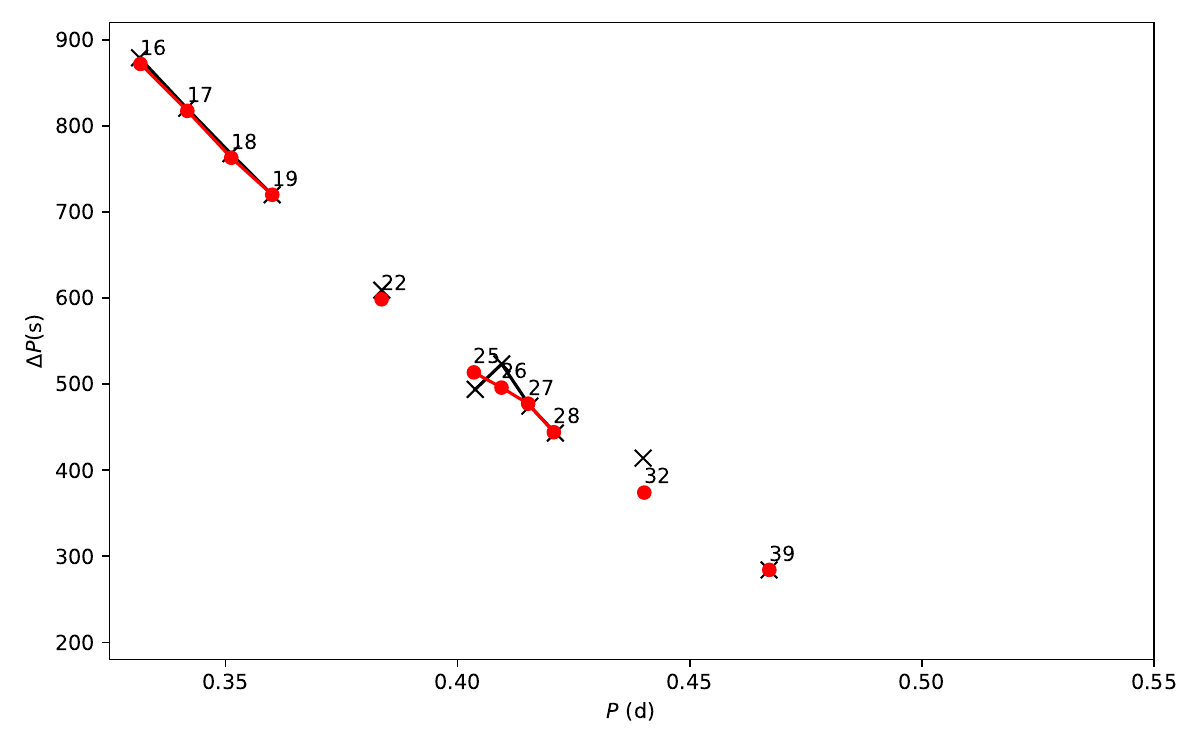}
\caption{\patpes}
\label{fig:bfcthreepatpestightr}
\end{subfigure}

\begin{subfigure}{0.5\textwidth}
\centering
\includegraphics[width=1\columnwidth]{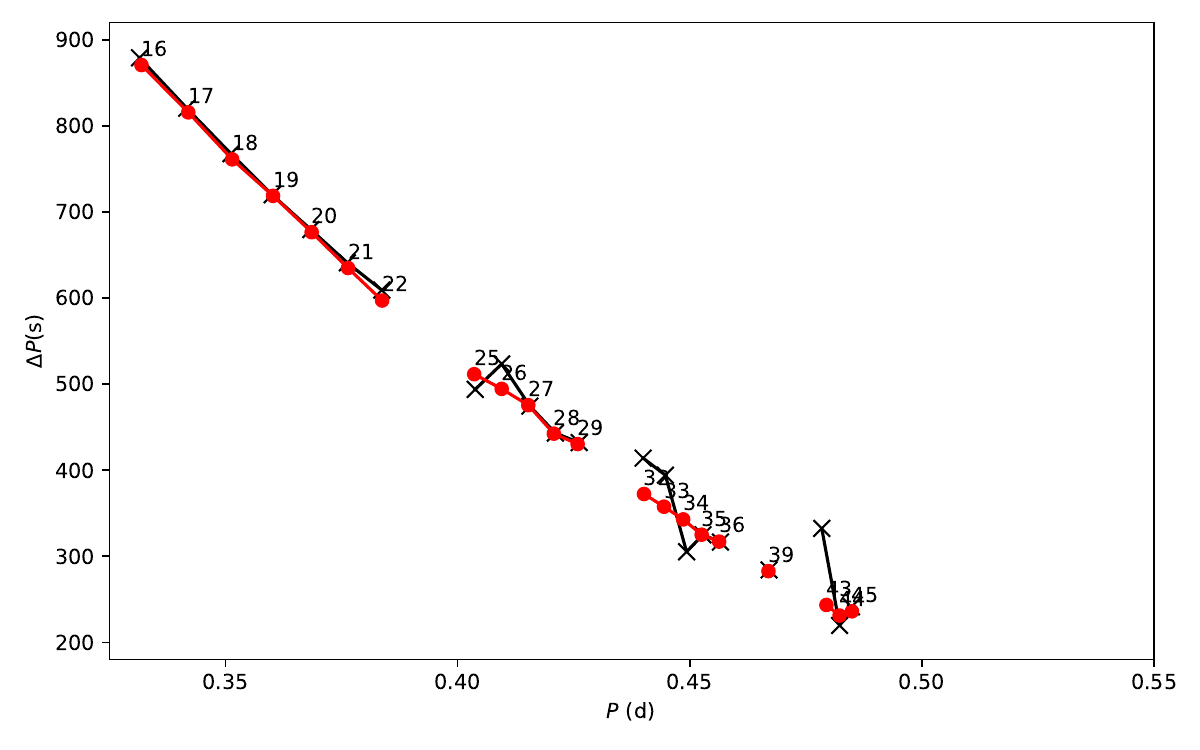}
\caption{\patopt}
\label{fig:bfcthreepatopttightr}
\end{subfigure}%
\begin{subfigure}{0.5\textwidth}
\centering
\includegraphics[width=1\columnwidth]{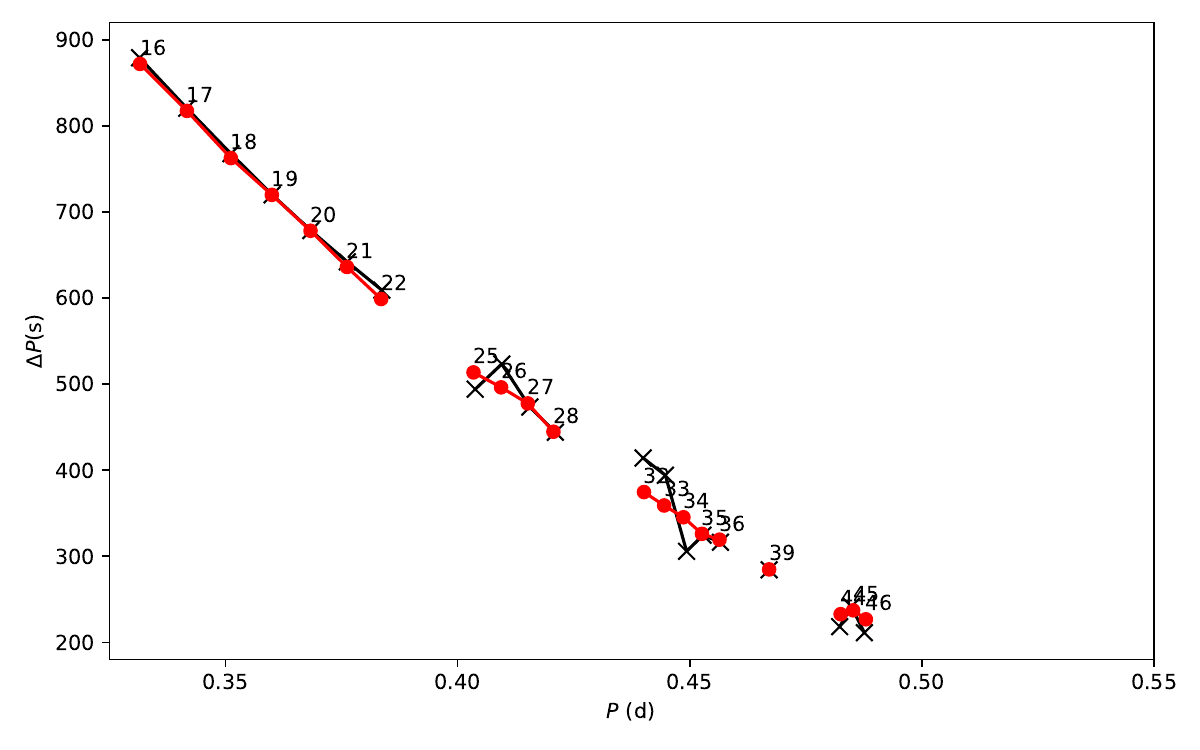}
\caption{\patsts}
\label{fig:bfcthreepatststightr}
\end{subfigure}
\caption{Best-fitting \cthree\ models (red) consistent with the tight  and spectroscopic constraints from the high resolution sampling. Observed patterns are shown in black.}
\label{fig:bfcthreetightr}
\end{figure*}

\begin{figure*}
\centering
   \includegraphics[width=2\columnwidth]{pics/highres/gangli/tightRgangli_monstrosity_chi2_hist_pie.pdf}
       \caption{\patli\ \patpes\ modelling results, tight radial constraint. See Fig. \ref{fig:monsterli} for detailed description of the different plot elements.}
    \label{fig:monsterlitightr}
\end{figure*}

\begin{figure*}
\centering
   \includegraphics[width=2\columnwidth]{pics/highres/conservative/tightRconservative_monstrosity_chi2_hist_pie.pdf}
       \caption{\patpes\ modelling results, tight radial constraint. See Fig. \ref{fig:monsterli} for detailed description of the different plot elements.}
    \label{fig:monsterpestightr}
\end{figure*}

\begin{figure*}
\centering
   \includegraphics[width=2\columnwidth]{pics/highres/hopeful/tightRhopeful_monstrosity_chi2_hist_pie.pdf}
       \caption{\patopt\ modelling results, tight radial constraint. See Fig. \ref{fig:monsterli} for detailed description of the different plot elements.}
    \label{fig:monsteropttightr}
\end{figure*}

\begin{figure*}
\centering
   \includegraphics[width=2\columnwidth]{pics/highres/sts/tightRsts_monstrosity_chi2_hist_pie.pdf}
       \caption{\patsts\ modelling results, tight radial constraint. See Fig. \ref{fig:monsterli} for detailed description of the different plot elements.}
    \label{fig:monsterststightr}
\end{figure*}

\section{Other frequency groups}
\label{sec:other_freq_groups}

The focus of this work has been asteroseismic exploitation of a dipole prograde gravito-inertial mode period-spacing pattern. However, there are some additional smaller groups of frequencies within the Fourier spectrum. Here, we briefly describe these features and speculate on their potential utility in the context of future, more refined modelling attempts.

\begin{figure*}
    \centering
    \includegraphics[width=0.9\linewidth]{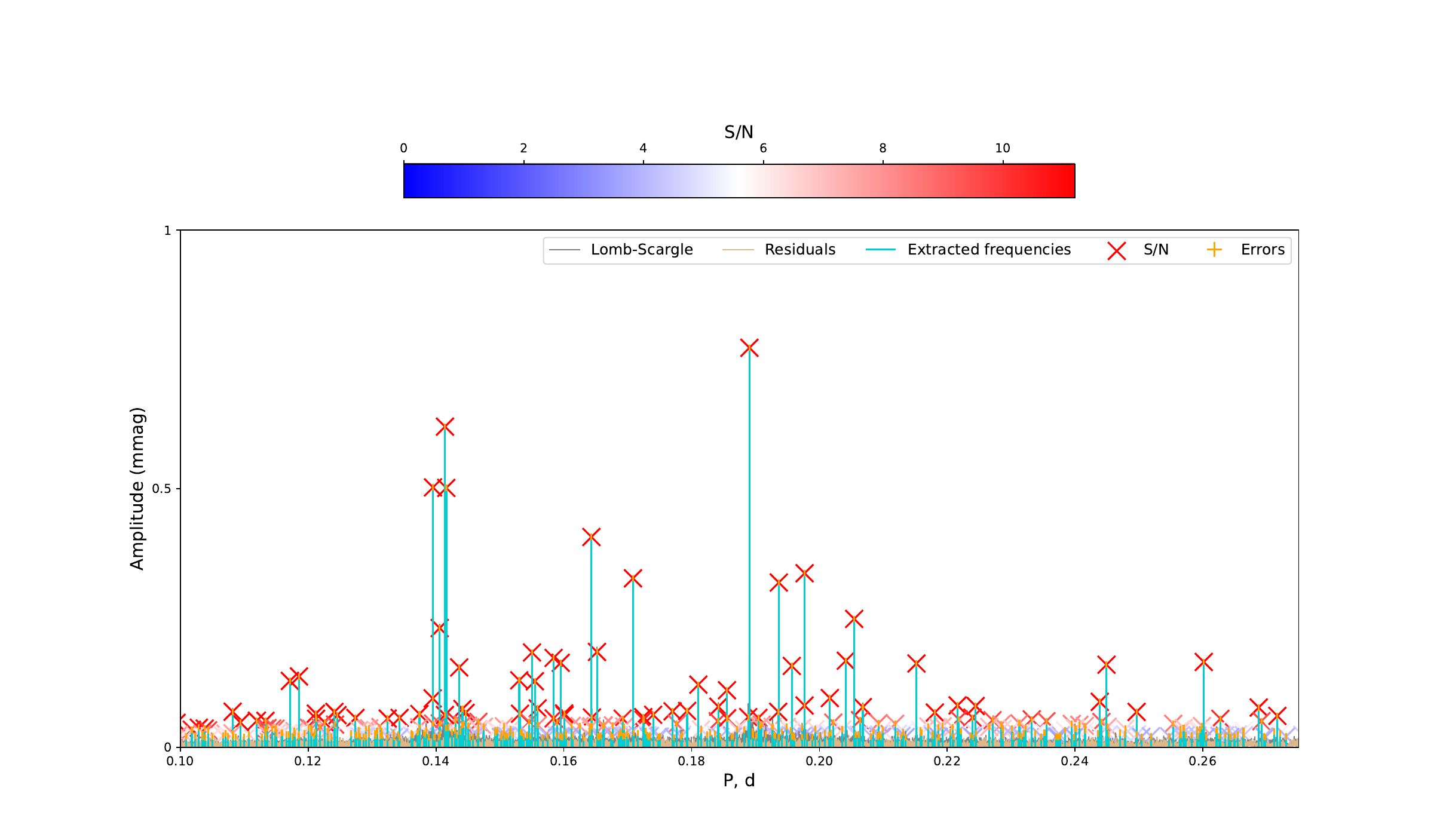}
    \caption{Likely high order g-modes, shown in the period domain. The cluster around 0.2~d are where we expect the l=2 series to appear, while those around 0.14 are likely the l=3 series. At most, a very short (perhaps 2-3 consecutive spacings) might be confidently obtained from the $l=2$ series.}
    \label{fig:high_gmode_series}
\end{figure*}

Between 0.1~d and 0.28~d in the periodogram, there are a large number of statistically significant periods where we expect quadrupole and septupole series to appear (around 0.14~d and 0.2~d, respectively). There is a great deal of contamination, however, and only a few consecutive periods in a (probable) quadrupole pattern can be identified. The relevant extracted frequencies are shown in Fig. \ref{fig:high_gmode_series}. These series did not feature as part of our analysis, the main component of which relied on asteroseismic models of prograde dipole modes.

\begin{figure*}
    \centering
    \includegraphics[width=0.9\linewidth]{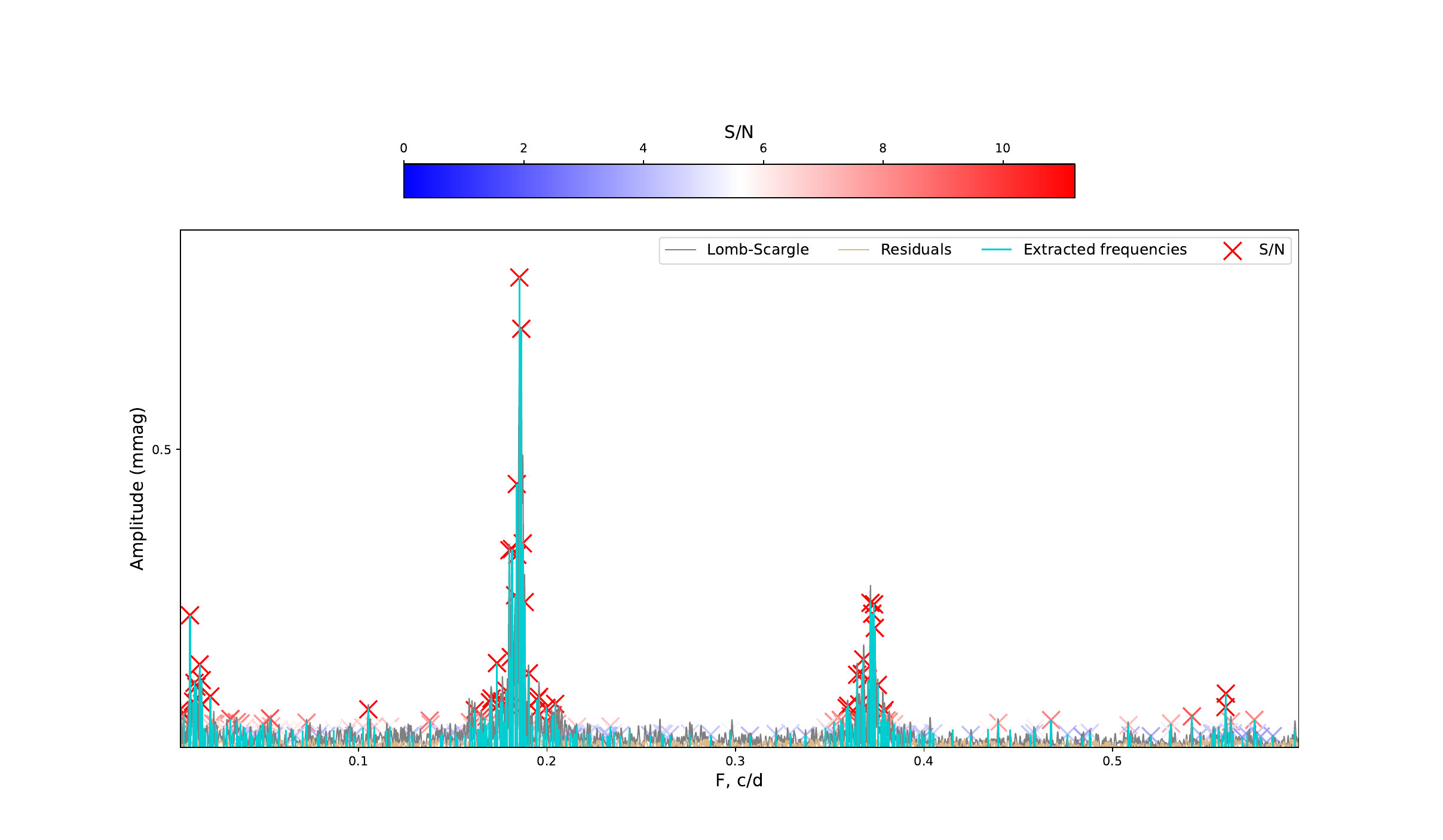}
    \caption{Series of low-frequency cones around 0.19, 0.38, and 0.56 \perd. Origin of these frequency cones is unknown, and does not appear to coincide with any of the existing periods or known rotation rates.}
    \label{fig:lowfreqcones}
\end{figure*}

As previously noted by \cite{heliminiak2017}, there are several low-frequency cones present in the Fourier spectrum around 0.19~\perd, 0.38~\perd, and 0.56~\perd, with the two higher frequency cones presumably being harmonics relating to the dominant 0.19~\perd\ cone. These cones are shown in Figure \ref{fig:lowfreqcones}.
\cite{heliminiak2017} propose that the 0.18~\perd\ is caused by rotation one or both of the two G stars Ba and Bb, commenting that the complicated structure of these peaks could be a result of either differential rotation or originate from two stars rotating at similar rates. From atmospheric modelling of the disentangled spectra in \cite{kemp2024eclipse}, the rotation rates for Ba and Bb were estimated to be $v_{\rm sin(i)}$=$8.9\pm1.1$ \kms\ and $v_{\rm sin(i)}$=$9.2\pm1.1$ \kms, respectively. From \cite{heliminiak2017}, we also have estimates for the stellar radii for the Ba and Bb of $0.888\pm0.010$R\solar\ and $0.856\pm0.038$R\solar, respectively. Assuming rigid rotation, this corresponds to rotation frequencies of approximately $0.2\pm0.02$\perd\ and $0.21\pm0.03$\perd. This is slightly higher than expected from the centre of the cone, although within uncertainty. It is likely that even a small systematic error in either the rotation velocity or the radius estimate could resolve this. We note that in the scenario of differential rotation, an offset between the surface rotation and the interior might be expected. However, according to our surface rotation estimate the surface would be rotating slightly faster than the interior, and significantly faster than the orbital frequency of the eccentric B binary (0.116~\perd). We conclude that the rotation of the B stars is almost certainly the cause of these cones, noting that the surface rotation rate is consistent with the cone frequencies to within $1-\sigma$.

Finally, there are the high-amplitude p-modes at 17.75~\perd, 18.48~\perd, 20.24~\perd, and 22.62~\perd (see Fig. \ref{fig:scg_sts}). They exhibit significant, symmetric splitting that is present after prewhitening. The splitting corresponds to the orbital frequency of the Aa binary, and may be due to light travel-time effects. Similar splittings are only occasionally visible in the g-mode regime, but are but are never statistically significant even for the high-amplitude peaks. There is no sign of rotational splitting.

\end{document}